\documentclass[useAMS,usenatbib]{mn2e}
\usepackage{amstext}
\usepackage{amssymb}
\usepackage{amsmath, float}
\usepackage{graphicx}
\usepackage{txfonts}
\usepackage{amsmath,mathtools,revsymb,amssymb}
\usepackage{booktabs,multirow,calc,xspace,rotating}
\usepackage{verbatim}

\newcommand       \be           {\begin{equation}}
\newcommand       \ee           {\end{equation}}
\newcommand       \bea          {\begin{eqnarray}}
\newcommand       \eea          {\end{eqnarray}}

\newcommand       \M		{\,{\cal M_{\rm conv} }}

\newcommand       \grad		{\rm \gamma_{\rad}}

\newcommand       \ome	{\omega_{e}}
\newcommand       \omc	{\omega_{c}}
\newcommand       \om	{\omega}
\newcommand	      \N	{\mathcal{N}}

\renewcommand        \la       {\langle}
\newcommand        \ra       {\rangle}

\renewcommand{\vec}[1]{\boldsymbol{#1}}

\renewcommand{\ddot}{\vec{\cdot}}

\renewcommand{\grad}{\vec{\nabla}}
\renewcommand{\div}{ \grad \ddot }

\newcommand{\A}{\mathcal{A}}
\renewcommand{\M}{\mathcal{M}}
\newcommand       \Ai        {{\rm Ai}}
\newcommand       \Bi        {{\rm Bi}}

 \newcommand{\acknowledgments}{\begin{small}
    \section*{Acknowledgments}\end{small}}

\begin{document}

\title[IGW Excitation]{Internal Gravity Wave Excitation by Turbulent Convection}
\author[D. Lecoanet \& E. Quataert]{D. Lecoanet$^{1}$\thanks{E-mail: dlecoanet@berkeley.edu} \& E. Quataert$^{1}$\thanks{E-mail: eliot@berkeley.edu}  \\
  $^{1}$Astronomy Department and Theoretical Astrophysics
 Center, University of California, Berkeley, 601 Campbell Hall,
 Berkeley CA, 94720\\ }

\date{Accepted . Received ; in original form }
\pagerange{\pageref{firstpage}--\pageref{lastpage}} \pubyear{????}
\maketitle

\begin{abstract} We calculate the flux of internal gravity waves (IGWs) generated by turbulent convection in stars.  We solve for the IGW eigenfunctions analytically near the radiative-convective interface in a local, Boussinesq, and cartesian domain.  We consider both discontinuous and smooth transitions between the radiative and convective regions and derive Green's functions to solve for the IGWs in the radiative region.  We find that if the radiative-convective transition is smooth, the IGW flux depends on the exact form of the buoyancy frequency near the interface.  IGW excitation is most efficient for very smooth interfaces, which gives an upper bound on the IGW flux of $\sim F_{\rm conv} (d/H)$, where $F_{\rm conv}$ is the flux carried by the convective motions, $d$ is the width of the transition region, and $H$ is the pressure scale height.  This can be much larger than the standard result in the literature for a discontinuous radiative-convective transition, which gives a wave flux $\sim F_{\rm conv} \M$, where $\M$ is the convective Mach number.  However, in the smooth transition case, the most efficiently excited perturbations will break in the radiative zone.  The flux of IGWs which do not break and are able to propagate in the radiative region is at most $\sim F_{\rm conv} \M^{5/8} (d/H)^{3/8}$, larger than the discontinuous transition result by $(\M H/d)^{-3/8}$.  The transition region in the Sun is smooth for the energy-bearing waves; as a result, we predict that the IGW flux is a few to five times larger than previous estimates.  We discuss the implications of our results for several astrophysical applications, including IGW driven mass loss and the detectability of convectively excited IGWs in main sequence stars.  \end{abstract}  \label{firstpage}

\vspace{-0.7cm}
\section{Introduction}
\voffset=-2cm
\vspace{-0.1cm}

Internal gravity waves (hereafter, IGWs) are a class of waves in a stably stratified background in which buoyancy serves as a restoring force.  IGWs propagate in radiative zones in stars and can influence composition, angular momentum, and energy transport within stars.  IGWs could also be important diagnostics of stellar structure---the detection of standing IGWs ($g$-modes) has been a long-standing goal of helioseismology \citep{Sev76,Bro76}, as $g$-modes provide better information about the core of the Sun than the more easily observed global sound waves ($p$-modes) \citep[e.g.,][]{TC01}.  However, IGWs are evanescent in the convection zone, so their surface manifestation is expected to be small.

IGWs have been invoked to explain the observation that F-stars have a smaller than expected Li abundance \citep[e.g.,][]{TC98}.  \citet{GLS91}, hereafter GLS91, first suggested that mixing from IGWs could enhance diffusion of Li, leading to lower Li abundances.  \citet{CT05} invoke IGWs to explain both the Li abundances of solar-type stars and the rotation of the solar interior.  When propagating through a differentially rotating star, selective damping of modes can deposit the wave's angular momentum and modify the star's rotation profile \citep[e.g.,][]{KQ97, ZTM97,TKZ02}.  Note, however, that IGWs generally have an anti-diffusive effect, accentuating angular velocity gradients.  This anti-diffusive behavior leads to the quasi-biennial oscillation (QBO) in the Earth's atmosphere, and has been studied extensively by the atmospheric science community \citep{Bal01,FA03}.

Massive stars have convective cores surrounded by a radiative envelope.  \citet{QS12} suggested that extremely vigorous convection within the last $\sim$ year of a massive star's life could generate a super-Eddington IGW flux and drive significant mass loss.  Earlier in a massive star's life, the angular momentum carried by IGWs may generate substantial differential rotation, perhaps mirroring the QBO in the Earth's atmosphere \citep{RLL12}.

In some stars, IGWs are linearly unstable, driven by, e.g., the $\epsilon$ or $\kappa$ mechanisms \citep{Unn89}.  Even absent such linear driving, however, IGWs are thought to be generated by turbulent convection.  Although IGWs are evanescent in a convective region, they can be excited by Reynolds stresses or entropy fluctuations associated with the convection.  A related excitation mechanism is IGW generation by overshooting convective plumes which penetrate into the radiative region.  Numerical simulations of a radiative zone adjacent to a convection zone find efficient generation of IGWs \citep[e.g.,][]{RG05,MA07,BMT11}.    Although simulations reported in \citet{RG05} \& \citet{MA07} show power distributed over a wide range of frequencies and wavelengths, the power spectra in \citet{BMT11} exhibit ridges corresponding to discrete $g$-modes.\footnote{The simulations of \citet{RG05} and \citet{BMT11} solve the anelastic equations, which do not conserve energy \citep{BVZ12}.  This could potentially produce errors in the IGW amplitudes and/or power spectra.}  Simulations often require artificially high diffusivities in the radiative zone to maintain a strong convective flux, and thus IGWs are artificially strongly damped in the radiative zone.  This complicates estimating IGW fluxes or quantitatively studying the effects of IGWs on the stellar structure.

There have been several efforts to analytically estimate the flux of IGWs stochastically excited by turbulent convection.  These models are essential for determining the resulting efficiency of the mixing, angular momentum transport, or mass-loss produced by IGWs.  \citet[hereafter P81]{Pre81} and GLS91 match pressure perturbations in the convective region to pressure perturbations in the waves, whereas \citet[hereafter GK90]{GK90} and  \citet[hereafter B09]{Bel09} calculate eigenmodes and derive how their amplitudes change using an inhomogeneous wave equation.  P81, GLS91, and GK90 all model the convective region using mixing length theory, assuming a Kolmogorov turbulence spectrum.  B09 uses an energy spectrum calculated from a direct numerical simulation of the solar convection zone.  Each of these papers predicts a different IGW power spectrum.

In this paper, we calculate the IGW flux generated by turbulent convection and clarify the relationship between different predictions in the literature.  In Section~\ref{sec:background}, we state our assumptions regarding the background state, and describe some properties of IGWs.  Our main calculation is in Section~\ref{sec:wavegeneration}, where we introduce our formalism for calculating the IGW flux.  Our formalism relies on calculating a Green's function using the eigenmodes of the system (also discussed in P81).  We relate our method to GK90's in Appendix~\ref{sec:norm}.  In Section~\ref{sec:waveflux} we calculate the IGW flux and rms wave displacements for both smooth and discontinuous radiative-convective transitions.  Next, we show that our results for a discontinuous transition can be derived more heuristically using pressure balance arguments (Section~\ref{sec:pressure}); we also make detailed comparisons to previous results (Section~\ref{sec:pastresults}).  Finally, in Section~\ref{sec:conclusion} we conclude, show how our results increase the predicted IGW flux in stars, and discuss some implications of this increased wave flux.

\section{Background State and Perturbation Equations}\label{sec:background}

\begin{figure}
  \begin{center}
    \includegraphics[width=\linewidth]{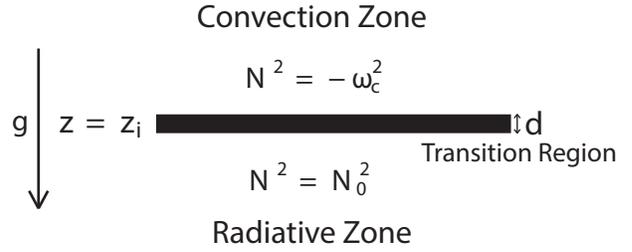}
  \end{center}
\caption{
A schematic of our problem setup.  The radiative-convective interface is at $z=z_i$, where $z_i$ is close to zero, and has width $d$.  Gravity points downward in the $z$ direction.  The convection zone is the region $z>z_i$ and the radiative zone is the region $z<z_i$.  We will use $\xi_{z,{\rm rad}}$ to denote the part of the vertical displacement within the radiative zone.  If $d$ is small, the waves see the radiative-convective transition as discontinuous; we will use superscript $D$ to denote results for a discontinuous transition.  If $d$ is large, the waves see the radiative-convective transition as smooth.  In this case, the results depend on the $N^2$ profile very close to $z_i$.  We consider $N^2$ parameterized by a $\tanh$ profile, which is a very smooth transition; we will use superscript $T$ to denote results for the $\tanh$ profile.  We also consider a piecewise linear $N^2$ profile, which is the most abrupt possible continuous transition; we will use superscript $L$ to denote results for the piecewise linear profile.  Eqns.~\ref{eqn:dfluxt},  \ref{eqn:sfluxr}, \& \ref{eqn:lfluxr} give our IGW flux estimates for discontinuous, $\tanh$, and piecewise-linear $N^2$, respectively.
  \label{fig:schematic}
}
\end{figure}

In this paper we consider a simple model of a radiative zone adjacent to a convection zone.  We assume that the length scales of interest are small in comparison to the stellar radius, i.e., we are in the local limit, so we use cartesian geometry, where $\vec{e}_z$ is the direction of gravity.  In our model, the radiative zone is the region $-L < z < z_i$, and the convection zone is the region $z_i < z < L$, where $z_i$ is the location of the radiative-convective interface, and both regions have a horizontal area $\A$.  We take $L$ and $\sqrt{\A}$ to be much larger than any other length scale in the problem, and will assume $z_i$ is close to zero.  In Figure~\ref{fig:schematic} we sketch a schematic of our model.  Using a domain with finite vertical extent provides simpler boundary conditions, but yields the same results as an infinite domain.

Furthermore, we employ the Boussinesq approximation.  This is appropriate if the wave generation occurs close to the radiative-convective boundary, and if we are only concerned with IGWs near this boundary.  We will see that the wave generation primarily occurs in a region with height approximately equal to the size of the energy bearing convective motions, which we assume is $\sim H$ the pressure scale height.  Although the Boussinesq approximation is only rigorously valid on length scales smaller than $H$, we recover results similar to those presented in GK90 who used the fully compressible equations.  We thus believe that our results would not change significantly if we used the fully compressible equations.

We model the radiative region as a stably stratified atmosphere with a squared buoyancy frequency $N_0^2$.  The convective region is much more complicated due to turbulent motions.  We decompose the fluid properties in the convection zone into time averaged and fluctuating components.  We assume the time averaged velocity is zero, and there is a very small mean stratification with squared buoyancy frequency $-\omc^2$.  Because the convective region is nearly adiabatic, $\omc\ll N_0$.  We treat the fluctuating components of the velocity and entropy in the convective region as source terms in the wave equation.  In practice, we only include source terms due to the Reynolds stress in our analysis; source terms due to entropy fluctuations are of the same size or smaller than the Reynolds stress terms (P81, GK90).

With these assumptions, the equation for the evolution of the vertical displacement $\xi_z$ is
\be\label{eqn:rad}
\grad^2\frac{\partial^2}{\partial t^2}\xi_z+N_0^2\grad_\perp^2\xi_z=0,
\ee
in the radiative region, and
\be\label{eqn:conv}
\grad^2\frac{\partial^2}{\partial t^2}\xi_z - \omc^2\grad_\perp^2\xi_z = S = -\grad^2 F_z+\frac{\partial}{\partial z}\grad\ddot\vec{F},
\ee
in the convective region.  We take $\grad_\perp=\partial_x\vec{e}_x+\partial_y\vec{e}_y$ to be the horizontal part of the gradient operator (perpendicular to gravity), and $S$ to be the source term due to the Reynolds stress $\vec{F}$.

There are three parts of the Reynolds stress $\vec{F}$ on the RHS of eqn.~\ref{eqn:conv}: the convection-convection term, $\div (\vec{u}_{\rm c}\vec{u}_{\rm c})$; the wave-convection terms, $\vec{u}_{\rm c}\ddot\grad\partial_t\vec{\xi} + (\partial_t\vec{\xi})\ddot\grad\vec{u}_{\rm c}$; and the wave-wave term, $(\partial_t\vec{\xi})\ddot\grad(\partial_t\vec{\xi})$.  In this paper, we will only consider the convection-convection term, taking
\be
\vec{F}=\div (\vec{u}_{\rm c}\vec{u}_{\rm c}).
\ee
Nonlinearities from the wave-wave term are only important if $k_z\xi_z\gtrsim 1$.  We will find later that this condition is not satisfied in the convection zone, although wave breaking does occur within the radiative zone.  The first wave-convection interaction term, $\vec{u}_{\rm c}\ddot\grad\partial_t\vec{\xi}$, is the advection of wave energy by convection, and thus does not change the wave energy.  The second part, $(\partial_t\vec{\xi})\ddot\grad\vec{u}_{\rm c}$, gives the effect of the strain associated with the convection on the wave, and can contribute to wave generation.  However, we find that the wave flux is smaller than the convective flux, so the wave velocities are smaller than the convective velocities.  Furthermore, one can check that the $(\partial_t\vec{\xi})\ddot\grad\vec{u}_{\rm c}$ is also smaller than the other linear (in $\vec{\xi}$) terms in our eigenvalue equation (e.g., using eqn.~\ref{eqn:amps} or eqn.~\ref{eqn:ampd}).  Thus, it is consistent to take $\vec{F}=\div(\vec{u}_{\rm c}\vec{u}_{\rm c})$.

We now discuss the wave solutions to the homogeneous equations, i.e., taking $S=0$.  Because the equations are autonomous in $x,y,t$, we can Fourier transform in these directions.  Thus, we can take the solutions to be
\be
\xi_z(x,y,z,t)=\xi_z(z)\exp(ik_x x+ik_y y-i\om t),
\ee
and define the horizontal wavenumber $k_\perp=\sqrt{k_x^2+k_y^2}$, i.e., the wavenumber perpendicular to gravity.  Throughout this paper we will assume $N_0\gg \om$.  The solutions to eqns.~\ref{eqn:rad},\ref{eqn:conv} are
\bea\label{eqn:radsol}
\xi_z  & = &  B_1\cos(N_0 k_\perp z/\om)+B_2\sin(N_0 k_\perp z/\om), \\
\xi_z  & = &  C_1\exp(-k_\perp z) + C_2\exp(k_\perp z), \label{eqn:convsol}
\eea
respectively, where we have assumed $\sqrt{\om^2+\omc^2}\sim\om$.  The horizontal displacement $\vec{\xi}_\perp$ and pressure perturbation $\delta p$ are related to $\xi_z$ by
\bea
\xi_\perp  & \sim &  i(N_0/\om) \xi_z, \label{eqn:radpol} \\
\delta p  & \sim &  i\rho_0(N_0\om /k_\perp) \xi_z,\label{eqn:raddp}
\eea
in the radiative region, and
\bea
\xi_\perp  & \sim &  \xi_z, \\
\delta p  & \sim &  \rho_0(\om^2/k_\perp) \xi_z, \label{eqn:convpol}
\eea
in the convective region.  The background density is $\rho_0$, which is constant to lowest order in the Boussinesq approximation.

To solve for the coefficients in eqns.~\ref{eqn:radsol} \& \ref{eqn:convsol} and the eigenvalues $\om$, we must impose four boundary conditions and a normalization condition (the latter is discussed in Section~\ref{sec:amplitude}).  Two of the boundary conditions are on the behavior of $\xi_z$ at $z = \pm L$.  The physical solution requires that $\xi_z=0$ at the top and bottom boundaries.  The other two boundary conditions are set at the radiative-convective interface, $z=z_i$.  These depend on the nature of the boundary between the radiative and convective regions, and determine which $\om$ satisfy the eigenvalue problem.  Assume that $N^2$ varies from $N_0^2$ to $-\omc^2$ in a thin layer with height $d$, as illustrated in Figure~\ref{fig:schematic}.  If there is a sharp transition between the radiative and convective regions, i.e., $(k_\perp N_0/\om)d \ll1$, we can make the approximation that $N^2$ is discontinuous at $z_i$, which we take to be at $z=0$.  However, if $N^2$ varies slowly, i.e., $(k_\perp N_0/\om) d\gg 1$, then interesting behavior can take place in the transition region.  As we discuss in Section~\ref{sec:conclusion}, we expect the most efficiently excited waves in the Sun to fall under this latter regime.  We will discuss both the discontinuous and smooth $N^2$ limits below.

\section{Wave Generation by Turbulent Convection}\label{sec:wavegeneration}

Because the wave generation and wave propagation regions are distinct, we use a Green's function (or equivalently, variation of parameters), as in P81.  Once we have a Green's function $G(z,t;\zeta,\tau)$, we can write the vertical displacement in the radiative region as
\be
\xi_{z,{\rm rad}} = \int_{\ -\infty}^{\ t} d\tau \int_{\ z_i}^{\ L} d\zeta \ G(z,t;\zeta,\tau) \ S(x,y,\zeta,\tau),
\label{eqn:greensfunction}
\ee
where we assume that $\xi_{z,{\rm rad}}$ is zero at $t\rightarrow -\infty$.  The Green's function depends on whether $N^2$ can be modeled as discontinuous or smooth at the radiative-convective boundary.  In Section~\ref{sec:discontinuous} we calculate the Green's function assuming $N^2$ is discontinuous (as was assumed in GLS91 and GK90) and then in Section~\ref{sec:continuous} we treat the smooth $N^2$ case.  As we shall argue, the latter is more appropriate for the low frequency waves which dominate the IGW flux.  In Appendix~\ref{sec:norm} we show that the Green's function method is formally equivalent to GK90's method of expanding $\xi_z$ into normal modes to solve eqns.~\ref{eqn:rad}, \ref{eqn:conv}.

\subsection{Green's Function for Discontinuous $N^2$}\label{sec:discontinuous}

To calculate the Green's function, we need two linearly independent solutions, one which satisfies $\xi_z(-L)=0$, and one that satisfies $\xi_z(+L)=0$.  The boundary conditions at $z_i$, which we take to be at $z=0$, when $N^2$ is discontinuous, are that $\xi_z$ and $\delta p$ are continuous at $z=0$.  The first solution, which we call $\eta^D_z$, satisfies the boundary condition at $z=+L$:
\be\label{eqn:eigd1}
\eta_z^D=\left\{
	\begin{array}{ll}
		B_1\cos( N_0 k_\perp z/\omega +\omega/N_0)  &  z < 0, \\
		B_1\exp(-k_\perp z) & z > 0.
	\end{array}
\right.
\ee
Here we use {\it superscript $D$} to denote the eigenfunction when $N^2$ is {\it discontinuous} at the interface.  Below, we will use  {\it superscript $T$} to denote quantities for a smooth $N^2$ parameterized by a $\tanh$ profile, and {\it superscript $L$} to denote quantities for a smooth piecewise linear $N^2$.  The second linearly independent solution, which we call $\xi_z^D$, satisfies the boundary condition at $z=-L$:
\be\label{eqn:eigd2}
\xi_z^D=\left\{
	\begin{array}{ll}
		B_2\sin( N_0 k_\perp z/\om)  &  z < 0, \\
		B_2\frac{N_0}{2\om}\left(\exp(k_\perp z)-\exp(-k_\perp z)\right) &  z > 0.
	\end{array}
\right.
\ee
The eigenvalues $\om$ must satisfy $\sin( N_0 k_\perp L/\om)=0$.  Later we will project the total vertical displacement in the radiative zone onto the basis $\{\xi_z\}_\om$.  The vertical displacement in the radiative zone is approximately orthogonal to $\{\eta_z\}_\om$ in the radiative zone.  Thus, it is important that our second linearly independent solution is also approximately orthogonal to $\{\eta_z\}_\om$, as is the case for eqns.~\ref{eqn:eigd1} \& \ref{eqn:eigd2}.

The general expression for the Green's function, assuming $z<\zeta$, is
\be\label{eqn:gengreen}
G(z,t;\zeta,\tau)=\int d{\om'} \ \frac{\delta(f(\om'))}{\om'^2} \frac{\xi_z(z;\om')\eta_z(\zeta;\om')}{W(\zeta)}\exp(-i\om'(t-\tau)),
\ee
where we label the eigenfunctions with their frequency $\om'$, $\delta$ denotes the Dirac delta function, and $W(\zeta)$ denotes the Wronskian of $\xi_z$ and $\eta_z$.  $f(\om')$ is a function which is zero if and only if $\om'$ is an eigenvalue.  For the discontinuous case, we have $f(\om')=\sin(N_0 k_\perp L/\om')$.  We thus can simplify eqn.~\ref{eqn:gengreen} to
\be
G(z,t;\zeta,\tau)=\sum_{\om'} \frac{1}{N_0 k_\perp L} \frac{\xi_z(z;\om')\eta_z(\zeta;\om')}{W(\zeta)}\exp(-i\om'(t-\tau)),
\ee
where the sum is over the eigenvalues $\om'$.  For the discontinuous $N^2$ problem, assuming $z<0$ and $\zeta>0$, the Green's function is
\be\label{eqn:disgreen}
G^D(z,t;\zeta,\tau) = \sum_{\om'} \ \frac{\om'}{N_0^2 k_\perp^2 L}\frac{1}{B_2} \xi_z^D(z;\om')\exp(-k_\perp \zeta-i\om' (t-\tau)).
\ee

\subsection{Green's Function for Smooth $N^2$}\label{sec:continuous}

If $N^2$ varies smoothly from $N_0^2$ to $-\omc^2$, then a WKB type approximation can be used, provided that $ N_0 k_\perp d/\om \gg 1$.  Our motivation for studying this limit is that the largest scale waves in stars satisfy $N_0 k_\perp d/\om \gg 1$ (see Section~\ref{sec:conclusion}).  We would like to develop an approximate solution which is valid within the transition region, allowing us to connect the solution in the radiative region (eqn.~\ref{eqn:radsol}) to the solution in the convective region (eqn.~\ref{eqn:convsol}).

The solution in the transition region depends on the form of $N^2(z)$ near the radiative-convective interface.  In this section, we will provide the some details of the calculation for a $\tanh$ profile.  An eigenmode with frequency $\om$ transitions from oscillatory behavior to exponential behavior at a point $z_t$ (where $N^2=\om^2$), which is lower than the the radiative-convective interface, $z_i$ (where $N^2=0$).  For a $\tanh$ profile, $z_t$ does not change very much as $\om$ changes; although it is smooth, it is not {\it too} smooth.  Thus, we believe that the $\tanh$ profile is the smoothest physically relevant $N^2$ profile.

In Appendix~\ref{sec:linear}, we also consider a piecewise linear $N^2$ profile.  In contrast to the smooth $\tanh$ profile, this is the most abrupt continuous transition possible.  Thus, we believe that any actual stellar $N^2$ profile should lie somewhere between these two limits.  Although we focus on the $\tanh$ profile in this section, we will also describe the IGW fluxes for the piecewise linear $N^2$ profile in Section~\ref{sec:waveflux}.

One might be tempted to appeal to WKB analysis to solve for the eigenfunction on either side of the interface, and then match across the interface by expanding $N^2$ to linear order near the wave's turning point (as is standard in, e.g., quantum mechanics).  Roughly, a WKB solution is valid if the local wavelength of the eigenfunction is small compared to the scale on which the wavenumber of the eigenfunction varies, which for us is $d$.  For smooth $N^2$ we have assumed $N_0 k_\perp d/\om\gg 1$, so the WKB solution in the radiative zone is valid.  However, the WKB solution might break down near the convection zone if $k_\perp d\ll 1$.  For the piecewise linear $N^2$ profile, a version of WKB matching is valid (see Appendix~\ref{sec:linear}).  For the $\tanh$ profile, however, the eigenfunction is poorly approximated by the WKB solution when $k_\perp d\ll 1$.  Moreover, because $d/H\leq 1$ and IGWs with $k_\perp H\sim 1$ dominate the wave flux (see Section~\ref{sec:waveflux}; $H$ here is the pressure scale height which we assume is the largest scale of the turbulence), the WKB solution fails for the most efficiently excited IGWs.  Instead, we need to develop a different method to solve for the eigenfunctions.  The details of this calculation are given in Appendix~\ref{sec:tanh}.

We assume $N^2(z)$ is given by
\be
N^2(z)= \frac{N_0^2+\omc^2}{2}\left(\tanh\left(-\frac{z}{d}\right)+1\right) -\omc^2.
\ee
In Appendix~\ref{sec:tanh}, we derive approximate forms for two independent eigenfunctions, and show that there is excellent agreement between the numerical solutions to the eigenvalue problem and our asymptotic Bessel function solutions.  We are interested in the behavior of the eigenfunctions near the radiative-convective interface $z_i$.  The interface is at
\be
\exp\left(-2\frac{z_i}{d}\right)\sim \frac{\omc^2}{N_0^2+\omc^2}.
\ee

The two independent solutions are
\bea\label{eqn:ssoln1}
\eta_z^{T}\sim \left\{
	\begin{array}{ll}
		B_1\cos( N_0 k_\perp z/\omega +\pi/4)  & z\gtrsim -d \\
		B_1\left(\frac{N_0k_\perp d}{\om}\right)^{1/2}\left(\frac{\om}{\bar{\om}}\right)^{k_\perp d} J_{\bar{d}}\left[\frac{\omc k_\perp d}{\om} \exp\left(-\frac{z-z_i}{d}\right)\right] & z\lesssim d
	\end{array} \right. \\ \label{eqn:ssoln2}
\xi_z^{T}\sim\left\{
	\begin{array}{ll}
		B_2\sin( N_0 k_\perp z/\om + \pi/4)  & z \gtrsim -d \\
		B_2\left(\frac{N_0k_\perp d}{\om}\right)^{1/2}\left(\frac{\bar{\om}}{\om}\right)^{k_\perp d}Y_{\bar{d}}\left[\frac{\omc k_\perp d}{\om} \exp\left(-\frac{z-z_i}{d}\right)\right] & z \lesssim d
	\end{array}
\right.
\eea
where $\bar{\om}^2=\omc^2+\om^2$ and $\bar{\om}/\om$ ranges between $\sqrt{2}$ and $1$ for $\omega \gtrsim \omega_c$, and $\bar{d}=\bar{\om}k_\perp d/\om$.  In eqns.~\ref{eqn:ssoln1} \& \ref{eqn:ssoln2} we have dropped several factors of order unity from the equations derived in Appendix~\ref{sec:tanh}.  The eigenvalues for this problem are the frequencies $\om$ which satisfy $\sin(- N_0 k_\perp L/\om +\pi/4)=0$.  In Figure~\ref{fig:eigenfunction} we plot $\eta_z^T$ for parameters representative of the energy-bearing waves in the sun.

Given eqns.~\ref{eqn:ssoln1} \& \ref{eqn:ssoln2}, the Green's function, for $z<0$ and $\zeta>0$ is
\bea
G^T(z,t;\zeta,\tau)\sim \sum_{\om'}  \left(\frac{\om' d}{N_0 k_\perp}\right)^{1/2} \left(\frac{\om'}{\bar{\om'}}\right)^{k_\perp d} \left(B_2 N_0 k_\perp L\right)^{-1} \nonumber \\
\times \ J \ \xi_z^T(z;\om')\exp(-k_\perp\zeta-i\om'(t-\tau)), \label{eqn:smoothgreen}
\eea
where we introduce the shorthand
\be
J\equiv J_{\bar{\om}k_\perp d / \om}\left(\frac{\omc k_\perp d}{\om}\right).
\ee
Using series expansions from \citet{AS72}, we can approximate
\be\label{eqn:besselapprox}
J\sim\left\{
\begin{array}{ll}
	1 & \mbox{if } k_\perp d \ll 1, \\
		(k_\perp d)^{-1/2}\exp(-k_\perp d)  & \mbox{if } k_\perp d \gg 1.
	\end{array}
\right.
\ee
Note that although the Green's function for a $\tanh$ profile is equal to the discontinuous Green's function when $N_0 k_\perp d/\om\sim 1$, the Green's function in eqn.~\ref{eqn:smoothgreen} is no longer valid when $N_0 k_\perp d/\om\ll 1$ (see Appendix~\ref{sec:numeric}).  Instead, eqn.~\ref{eqn:disgreen} must be used in this limit.

\subsection{Amplitude Equation}\label{sec:amplitude}

Now that we have the Green's function, we can calculate mode excitation.  First, we will expand $\xi_{z,{\rm rad}}$ (in eqn.~\ref{eqn:greensfunction}) into eigenmodes $\xi_{z,{\rm rad}}(z;\om)$.  We use the subscript ${\rm rad}$ to denote the $z<z_i$ part of the eigenfunctions (eqns.~\ref{eqn:eigd2}, \ref{eqn:ssoln2}).  We write
\be
\xi_{z,{\rm rad}}=\frac{1}{\sqrt{\A}}\sum_{\om'} \ A(t;\om') \ \xi_{z,{\rm rad}}(z;\om') \exp(ik_x x+ik_y y -i\om' t),
\ee
where the $\xi_{z,{\rm rad}}$ are the $z<z_i$ part of the eigenmodes. Using this representation in eqn.~\ref{eqn:greensfunction}, we take the inner product with $\xi_{z,{\rm rad}}(z;\om)$, multiply by $\exp(-ik_x x-ik_y y +i\om t)$, and integrate over $dxdy$ to find
\bea
A(t;\om) = \frac{1}{\sqrt{\A}} \int_{\ -\infty}^{\ t} d\tau \int dxdy \int_{\ z_i}^{\ L}d\zeta \frac{1}{N_0 k_\perp L}\frac{\eta_z(\zeta;\om)}{W(\zeta)} \nonumber \\
\times \ S(x,y,\zeta,\tau) \exp(-ik_xx-ik_yy+i\om\tau). \label{eqn:genamp}
\eea
This procedure is discussed more thoroughly in Appendix~\ref{sec:norm}.

At this point we must pick a normalization condition for our eigenfunctions.  The energy in the perturbation is
\bea
&& \int d^3x\rho_0\left|\frac{\partial}{\partial t}\vec{\xi}_{\rm rad}(z)\right|^2= \nonumber \\
&& \sum_\om \sum_{\om'} \ A(\om) \ A^*(\om') \ \left(\om\om' \int dz \ \rho_0 \vec{\xi}_{\rm rad}(z,\om) \vec{\cdot} \vec{\xi}^*_{\rm rad}(z,\om')\right).
\eea
We want to identify $\sum_{\om} |A(\om)|^2$ with the energy, so our normalization condition is
\be\label{eqn:normalization}
\om \om'\int dz \ \rho_0 \vec{\xi}_{\rm rad}(z;\om) \vec{\cdot} \vec{\xi}^*_{\rm rad}(z;\om')=\delta_{\om\om'},
\ee
where $\delta$ is the Kronecker delta.  Using the eigenfunctions (eqns.~\ref{eqn:eigd1}, \ref{eqn:eigd2}, \ref{eqn:ssoln1}, \ref{eqn:ssoln2}) and the polarization relation (eqn.~\ref{eqn:radpol}), the normalization condition implies
\be\label{eqn:B}
B^2 \sim B_1^2 \sim B_2^2 \sim \frac{1}{N_0^2 L \rho_0},
\ee
for all the $N^2$ profiles considered in this paper.  Using this normalization in eqn.~\ref{eqn:genamp}, the amplitude equations are
\bea
& & A^D(t;\om) = \frac{1}{\sqrt{\A}}\int_{\ -\infty}^{\ t} d\tau \int dxdy \ \exp(-i k_x x - i k_y y + i\omega \tau) \nonumber \\
& & \times \ \frac{\om }{ N_0 k_\perp^2 } \sqrt{\frac{\rho_0}{L}}\int_{\ z_i}^{\ L} d\zeta \exp(-k_\perp\zeta) \ S(x,y,\zeta,\tau), \label{eqn:amps} \\
& & A^T(t;\om) = \frac{1}{\sqrt{\A}}\int_{\ -\infty}^{\ t} d\tau \int dxdy \exp(-ik_x x-ik_y y+i\om \tau) \nonumber \\
& & \times \ \left( \frac{J \sqrt{\om \rho_0 d}}{\sqrt{ N_0 L k_\perp^3}}\right) \left(\frac{\om}{\bar{\om}}\right)^{k_\perp d} \int_{\ z_i}^{\ L} d\zeta \exp(-k_\perp \zeta) S(x,y,\zeta,\tau). \label{eqn:ampd}
\eea
It is straightforward to derive the analogous amplitude equation for the piecewise linear $N^2$ profile using the Green's function given in eqn.~\ref{eqn:linGF}.

\subsection{Model of Turbulent Convection}\label{sec:conv}

To make further progress, we need to specify the source term $S$.  We assume that the convective turbulence is composed of a large number of incoherent eddies, estimate the wave generation due to a single eddy in isolation, and then find the total wave generation by summing over all eddies. We model the statistical properties of stellar convection using Kolmogorov turbulence (see, e.g., \citealt{gk77}):  the  convective velocity on the outer-scale $H$ is $u_c$ and the associated convective turnover frequency is $\omega_c \sim u_c/H$.   The convective energy flux is $F_{\rm conv} \sim \rho_0 u_c^3$.
On scales $h$ sufficiently small compared to $H$, the turbulent power-spectrum is given by the Kolmogorov scaling:
\be u_h \simeq u_c \, (h/H)^{1/3} \simeq u_c (\ome/\omega_c)^{-1/2}
\label{vc}
\ee
where we have used the fact that smaller eddies have higher frequencies, i.e., shorter turnover times, with $\ome \simeq  u_h/h \propto h^{-2/3}$ and thus $h \propto \ome^{-3/2}$.   A given convective eddy characterized by its frequency $\omega_e$ can excite waves having frequencies $\omega$ and horizontal wavenumbers $k_\perp$ that satisfy
\be
\omega \lesssim \omega_e \ \ \ {\rm and}  \ \ \  k_\perp \lesssim k_\perp^{max} \simeq H^{-1} (\ome/\omc)^{3/2}.
\label{omkh}
\ee

\subsection{Energy Generation Rates and IGW Fluxes}\label{sec:waveflux}

In this section we calculate the IGW fluxes for discontinuous, tanh, and piecewise linear convective-radiative transitions.  We begin by estimating the energy generation due to a single eddy with size $h$ and turnover frequency $\ome$.  The source term contains three spatial derivatives which we can integrate by parts.  The contribution due to the source term is
\be
S\sim k_\perp^3 u_h^2.
\ee
Assuming the eddy has volume $h^3$ and lasts for a time $\ome^{-1}$, we can estimate the change in the amplitude due to a single eddy
\bea
& & \Delta A^D(\om) \sim \sqrt{\frac{\rho_0}{\A L}} \frac{\om}{N_0} k_\perp h^4 u_h, \\
& & \Delta A^T(\om) \sim \sqrt{\frac{N_0}{\om}} \left(J \sqrt{k_\perp d}\right)\left(\frac{\om}{\bar{\om}}\right)^{k_\perp d} \Delta A^D(\om), \\
& & \Delta A^L(\om) \sim \left(\frac{N_0 k_\perp d}{\om}\right)^{1/6} \Delta A^D(\om).
\eea

The total energy generation rate due to all eddies is then
\bea \label{eqn:Egend}
& & \dot E^D(\om) \sim \frac{\left(\Delta A^D\right)^2}{\ome^{-1}} \ \left(\frac{\A k_\perp^{-1}}{h^3}\right) \sim \frac{\rho_0}{L} \left(\frac{\om}{N_0}\right)^2 u_h^3 h^3 \ (k_\perp h), \\
& & \dot E^T(\om) \sim \frac{N_0}{\om} \left(J^2 k_\perp d\right)\left(\frac{\om}{\bar{\om}}\right)^{2k_\perp d} \dot E^D(\om), \\
& & \dot E^L(\om) \sim \left(\frac{N_0 k_\perp d}{\om}\right)^{1/3} \dot E^D(\om).
\eea
The factor of $\A k_\perp^{-1}/h^3$ in eqn.~\ref{eqn:Egend} counts the number of eddies with size $h$ which excite IGWs with frequency $\om$.  We have assumed excitation happens in a region with thickness $dz \sim k_\perp^{-1}$ (because the IGW eigenfunction decreases in the convection zone over a characteristic lengthscale $\sim k_\perp^{-1}$).  Because of the random phases of the convective eddies, the excitations due to different eddies are assumed to be uncorrelated, and the energy increases only linearly with the number of eddies.

In the case of smooth $N^2$, the flux decreases exponentially for $k_\perp d\gg 1$.  The dominant contribution to the flux is from $k_\perp d \lesssim 1$, so for the rest of this section, we will assume $k_\perp d\lesssim 1$.  The IGW flux is then given by
\bea
& & \frac{d F^D}{d\log\om \ d\log k_\perp} \sim \frac{ \dot E^D(\om)}{\A} \ \left(\A k_\perp^2 L k_\perp\frac{ N_0}{\om} \right) \nonumber \\
& &\sim \rho_0 u_h^3 \frac{\om}{N_0} (k_\perp h)^4 \sim \ \rho_0 u_c^3 \M (k_\perp H)^4 \left(\frac{\om}{\omc}\right)^{-13/2}, \label{eqn:dflux} \\
& & \frac{d F^T}{d\log\om \ d\log k_\perp} \sim \rho_0u_c^3 (k_\perp H)^4\left(\frac{\om}{\omc}\right)^{-15/2} \left(k_\perp d\right), \label{eqn:sflux} \\
& & \frac{d F^L}{d\log\om \ d\log k_\perp} \sim \rho_0 u_c^3 \M^{2/3} (k_\perp H)^4 \left(\frac{\om}{\omc}\right)^{-41/6} (k_\perp d)^{1/3}. \label{eqn:lflux}
\eea
where $\M=\omc/N_0$ is the convective Mach number.  The term in parentheses in the first equality of eqn.~\ref{eqn:dflux} is the density of states.  There are $\A k_\perp^2$ modes in the horizontal direction, and $Lk_\perp N_0/\om$ modes in the vertical direction, with wavenumber $\sim k_\perp$ and frequency $\sim \om$, which each contribute a flux $\dot E(\om)/\A$.  Recall that eqns.~\ref{eqn:dflux}-\ref{eqn:lflux} only apply for $\om \gtrsim \omc$ and $k_\perp \lesssim k_\perp^{\rm max}(\om)\sim H^{-1} (\om/\omc)^{3/2}$, and eqns.~\ref{eqn:sflux} \& \ref{eqn:lflux} assume $k_\perp d\lesssim 1$.

Integrating over $k_\perp$, we find
\bea
& & \frac{d F^D}{d\log\om} \sim  \rho_0u_c^3 \M \left(\frac{\om}{\omc}\right)^{-1/2}, \\
& & \frac{d F^T}{d\log\om} \sim \rho_0u_c^3 \left(\frac{d}{H}\right), \\
& & \frac{d F^L}{d\log\om} \sim \rho_0u_c^3 \M^{2/3} \left(\frac{\om}{\omc}\right)^{-1/3} \left(\frac{d}{H}\right)^{1/3}.
\eea
Finally, we find that the total flux is
\bea\label{eqn:dfluxt}
&& F^D\sim \rho_0u_c^3 \M\sim F_{\rm conv} \M, \\
&& F^T\sim \rho_0u_c^3 \left(\frac{d}{H}\right)\sim F_{\rm conv} \left(\frac{d}{H}\right), \label{eqn:fluxtt} \\
&& F^L\sim\rho_0u_c^3 \M^{2/3} \left(\frac{d}{H}\right)^{1/3} \sim F_{\rm conv} \M^{2/3} \left(\frac{d}{H}\right)^{1/3}. \label{eqn:dfluxl}
\eea

This estimate predicts, for a $\tanh$ $N^2$ profile, an IGW flux only slightly smaller than the convective flux.  However, as we now show, energy-bearing waves in the smooth $N^2$ case  (both tanh and piecewise linear profiles) will undergo vigorous wave-breaking within the radiative zone (see Figure~\ref{fig:eigenfunction}).  This process occurs concurrently with overshooting convective plumes, but is much more spatially localized (in $z$) than overshooting convection.

To quantify this argument, we calculate the typical size of the perturbations in the radiative zone using
\be
\frac{d F}{d\log\om \ d\log k_\perp}\sim \rho_0 (\om\xi_\perp)^2 u_{g,z},
\ee
where $u_{g,z}\sim \om/k_z \sim \om^2/(N_0 k_\perp)$ is the vertical group velocity, and we have assumed $\xi_\perp \gg\xi_z$ (eqn.~\ref{eqn:radpol}).  From this, we find
\bea\label{eqn:dxi}
&& \xi_z^D \sim H \frac{\om}{N_0} (k_\perp H)^{5/2} \left(\frac{\om}{\omc}\right)^{-21/4}, \\
&& \xi_z^T\sim H \sqrt{\frac{\om}{N_0}} (k_\perp H)^{3} \left(\frac{\om}{\omc}\right)^{-21/4} \left(\frac{d}{H}\right)^{1/2}, \\
&& \xi_z^L\sim H \left(\frac{\om}{N_0}\right)^{5/6} (k_\perp H)^{8/3} \left(\frac{\om}{\omc}\right)^{-21/4} \left(\frac{d}{H}\right)^{1/6},
\eea
and
\bea
&& k_z \xi_z^D\sim (k_\perp H)^{7/2}\left(\frac{\om}{\omc}\right)^{-21/4}, \\
&& k_z \xi_z^T\sim \M^{-1/2} (k_\perp H)^{4} \left(\frac{\om}{\omc}\right)^{-23/4} \left(\frac{d}{H}\right)^{1/2}, \\
&& k_z\xi_z^L\sim \M^{-1/6} (k_\perp H)^{11/3} \left(\frac{\om}{\omc}\right)^{-65/12}\left(\frac{d}{H}\right)^{1/6},
\eea
where we have used $k_z=k_\perp N_0/\om$ which holds in the radiative zone for $|z|\ll d$.  Recall that the condition for wave breaking is $k_z\xi_z\sim 1$.  For the case of discontinuous $N^2$, the most efficiently excited waves are marginally susceptible to wave breaking.  However, for both $\tanh$ and piecewise linear $N^2$, the most efficiently excited waves will break in the radiative zone.

The only waves that successfully propagate in the radiative zone have $k_z\xi_z \lesssim 1$.  Thus, to find the IGW flux for smooth $N^2$, we must integrate the flux only over the regions of $(k_\perp,\om)$ space in which $k_z\xi_z \lesssim 1$.  This implies
\bea
&& (H \M/d)^{1/2}\lesssim (k_\perp H)^4 \left(\frac{\om}{\omc}\right)^{-23/4} \ \ \ \ \ \mbox{($\tanh$)} \\
&& (H\M/d)^{1/6} \lesssim (k_\perp H)^{11/3}\left(\frac{\om}{\omc}\right)^{-65/12} \ \ \  \mbox{(piecewise linear)}
\eea
and, as before, $\om \gtrsim \omc$, $k_\perp\lesssim H^{-1} (\om/\omc)^{3/2}$, and $k_\perp d \lesssim 1$.  We find that the waves that are marginally susceptible to wave breaking, and which maximize the flux for the $\tanh$ profile are at the convective turnover frequency, $\om\sim\omc$, but have small wave numbers, $k_\perp H\sim (\M H/d)^{1/8}$.  For the piecewise linear profile, the spatial scale is $k_\perp H \sim 1$, but the waves have higher frequencies, $\om \sim \omc (\M H/d)^{-2/65}$.  The resulting IGW flux in waves that do not break is given by
\bea\label{eqn:sfluxr}
&& F^T\sim F_{\rm conv} \M^{5/8} \left(\frac{d}{H}\right)^{3/8}, \\
&& F^L\sim F_{\rm conv} \M^{57/65} \left(\frac{d}{H}\right)^{8/65}. \label{eqn:lfluxr}
\eea
These results are only valid if these waves see a smooth $N^2$ profile, i.e.,
\be
\M H /d \ll 1.
\ee
If this condition is satisfied, then the IGW flux is larger than that predicted by the discontinuous result by $(\M H/d)^{-3/8}$ for the $\tanh$ profile and $(\M H/d)^{8/65}$ for the piecewise linear profile.  Note that if $d/H\sim \M$, then the discontinuous and smooth $N^2$ limits give the same wave flux.

\subsection{Wave Excitation Within the Overshoot Region}
\label{sec:overshoot}

In the previous sections, we have consider the efficiency of IGW excitation by turbulent motions in the convection zone.  However, convective overshoot and wave breaking produce turbulent motions within the radiative zone, near the radiative-convective interface.  We can estimate the wave excitation within the radiative zone by convolving the Reynolds stress associated with turbulent motions due to convective overshoot with the appropriate Green's functions (see, e.g., Section~\ref{sec:continuous}).

The principal difficulty in calculating the wave generation in the overshoot region is in accurately describing the turbulent motions near the radiative-convective interface.  Although convective overshoot has been investigated via simulations \citep[e.g.,][]{RG052}, it is currently computationally infeasible to employ a realistic Mach number and interface stiffness.  To roughly estimate the IGW generation due to turbulent motions within the overshoot region, we will assume that the motions can be decomposed into incoherent eddies with the statistical properties of Kolmogorov turbulence, as above.  However, instead of taking the outer-scale of the cascade to be $H$, we will assume it is given by the size of the overshoot region $\sim d \log(N_0/\omc)$.  We assume the typical velocity on this outer-scale is still $u_c$.

\begin{figure}
  \begin{center}
    \includegraphics[width=\linewidth]{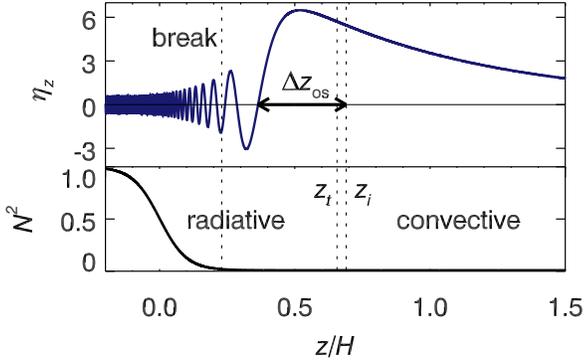}
  \end{center}
\caption{
Representative eigenfunction and buoyancy frequency squared near the radiative-convective transition.  See Appendix~\ref{sec:numeric} for details on the calculation of the eigenfunction.  The top panel shows the numerically calculated vertical perturbation eigenfunction normalized to have amplitude one in the radiative zone, for the parameters $\omc/N_0=10^{-3}$, $d/H=0.1$, and $k_\perp H=1$.  The bottom panel shows the buoyancy frequency squared normalized to one in the radiative zone, which we have assumed follows a tanh profile.  The vertical dotted lines, from left to right, correspond to the point at which $k_z\xi_z=1$ where we expect the mode to break; the transition point $z_t$ (defined by $N^2=\om^2$), where this mode transitions from exponential to oscillatory behavior, and gives a typical amplitude of $\eta_z$ within the overshoot region; and the radiative-convective interface $z_i$ (defined by $N^2=0$).  We have also labelled the distance between the radiative-convective interface and the first zero of the eigenfunction, $\Delta z_{\rm os}$.  Turbulent eddies associated with convective overshoot cannot efficiently couple to this mode unless they have vertical size less or equal to $\Delta z_{\rm os}$.
  \label{fig:eigenfunction}
}
\end{figure}

To predict where turbulent eddies can most effectively excite IGWs, it is helpful to consider the structure of the Green's function in the transition region.  In Figure~\ref{fig:eigenfunction}, we plot the eigenfunction $\eta_z^{T}(z)$ when $N^2$ is given by a tanh profile, with $\omc/N_0=10^{-3}$ and $d/H=0.1$, as we might expect for the energy-bearing eddies in the Sun (see Section~\ref{sec:conclusion}).

If an eddy is much larger than the local wavelength of the eigenfunction, then it will not be able to efficiently couple to the mode, as its convolution with the Green's function will to first order average out to zero.  The most efficient wave excitation in the overshoot region for the example mode in Figure~\ref{fig:eigenfunction} will be for eddies filling the region between $z_i$ and the first zero of $\eta_z$; we define this distance to be $\Delta z_{\rm os}$.  This eddy has size $\sim 0.3H$, smaller than the energy-bearing eddies of size $H$ in the convection zone.


Wave excitation in the overshoot region differs from wave excitation in the convection zone in several ways.  First, because we assume the outer-scale of the turbulence is $d\log(N_0/\omc)$ instead of $H$, the turbulent velocities on any length scale $h<H$ are larger in the turbulent velocities on that length scale in the convection zone (see Section~\ref{sec:conv}).  We will assume that the excitation within the overshoot region is given by eddies with size at most $\Delta z_{\rm os}$.  Thus, there are $\A \Delta z_{\rm os}/h^3$ eddies with size $h$ which excite IGWs with frequency $\om$ (see eqn.~\ref{eqn:Egend} and accompanying text).  Because the vertical length scale of $\eta_z$ is $\Delta z_{\rm os}$ in the overshoot region, the typical size of the Reynolds stress source term in eqn.~\ref{eqn:conv} is
\be
S\sim \frac{k k_\perp u_h^2}{\Delta z_{\rm os}},
\ee
where $k$ is the total wavenumber defined by
\be
k^2 = k_\perp^2 + \Delta z_{\rm os}^{-2}.
\ee
If $k_\perp\gg \Delta z_{\rm os}^{-1}$, then $k\approx k_\perp$, and if $\Delta z_{\rm os}^{-1}\gg k_\perp$ then $k\approx \Delta z_{\rm os}^{-1}$.  When we derived the Green's functions above (e.g., Section~\ref{sec:continuous}), we took $\eta_z(z_i)$ as a typical value of $\eta_z$ in the convection zone.  Here, we will take $\eta_z(z_t)$ as a typical value of $\eta_z$ in the overshoot region.

The exact form of the IGW wave flux depends on the background $N^2$ profile.  As an illustrative example, we will sketch the results for the tanh profile.  Broadly speaking, our estimates for wave excitation in the overshoot region are comparable to, but mostly smaller than, the wave excitation in the convection zone, except for high wavenumber waves with $k_\perp d\gg 1$ which are strongly suppressed in the convection zone.  Note that these results are predicated on our assumptions regarding the turbulence within the overshoot region, which are uncertain.  More detailed calculations likely require input from numerical simulations of plumes in the overshoot region.

For the tanh profile, the distance between the radiative-convective interface and the first zero of the eigenfunction, $\Delta z_{\rm os}$, is given by
\be
\Delta z_{\rm os} \approx d \log\left(\frac{\om}{\omc k_\perp d}\left(\frac{3\pi}{4}+\frac{\bar{\om}k_\perp d}{\om}\frac{\pi}{2}\right)\right).
\ee
We also have
\be
\eta_z(z_t)\sim \left(\frac{N_0k_\perp d}{\om}\right)^{1/2} \left(\frac{\om}{\bar{\om}}\right)^{\bar{\om}k_\perp d/\om}\overline{J},
\ee
where we use the shorthand
\be
\overline{J}=J_{\bar{\om}k_\perp d/\om}\left(\frac{\bar{\om}k_\perp d}{\om}\right)\sim\left\{
\begin{array}{ll}
	1 & \mbox{if } k_\perp d \ll 1, \\
		0.45 (k_\perp d)^{-1/3}  & \mbox{if } k_\perp d \gg 1.
	\end{array}
\right.
\ee
Note that $\overline{J}$ falls off much less steeply for $k_\perp d\gg 1$ than the associated convection zone quantity, $J$ (eqn.~\ref{eqn:besselapprox}).

Using these results, we can calculate the IGW power spectrum.  Because the result depends sensitively on our assumptions regarding the turbulence within the overshoot region, we will only highlight the general properties of the excitation power spectrum.  The IGW flux in the energy bearing mode, which has $k_\perp\sim H^{-1}$ and $\om\sim\omc$, is smaller in the overshoot region by a factor of $(\Delta z_{\rm os}/H)^2(H/(d\log(N_0/\omc))$; if we take $d/H\sim 0.1$ and $\omc/N_0=10^{-3}$ (see Section~\ref{sec:conclusion}), this factor is $\sim 0.2$.  For higher frequency waves with $\om/\omc=(H/\Delta z_{\rm os})^{3/2}$, the excitation is larger in the overshoot region by a factor of $(H/(d\log(N_0/\omc)),$ which is $\sim 1.5$ for the parameters given above.  Excitation is significantly more efficient in the overshoot region for modes with $k_\perp d\gg 1$.

As pointed out in Section~\ref{sec:waveflux}, there is a significant flux of IGWs which break in the radiative zone.  The breaking occurs where the local $k_z$ becomes comparable to $\xi_z^{-1}$.  As can be seen in Figure~\ref{fig:eigenfunction}, this occurs when $k_z$ is large (for higher frequency waves, the breaking would occur for even larger $k_z$).  If the turbulence associated with the wave breaking is isotropic, then only very small eddies would efficiently couple to the eigenfunction, leading to negligible wave excitation.  However, the wave will be very anisotropic when it breaks, possibly leading to more efficient wave excitation.  The details of wave generation by wave breaking are beyond the scope of this paper.

\section{Pressure Perturbation Balance}\label{sec:pressure}

A more heuristic way to derive the IGW flux is to compare the pressure perturbation on either side of the radiative-convective boundary.  This argument is not sufficiently precise to treat the smooth $N^2$ case---hence, we will assume $N^2$ is discontinuous, and thus that the pressure perturbation is continuous at the radiative-convective interface at $z=0$.  The pressure perturbation associated with a convective eddy with a turnover frequency $\ome$ and size $h$ is
\be\label{eqn:pconv}
\delta p_{\rm conv}\sim \rho_0 v_h^2 \sim \rho_0 u_c^2 (\ome/\omc)^{-1}.
\ee
The polarization condition (eqn.~\ref{eqn:raddp}) relates the pressure perturbation in the radiative zone to the vertical displacement,
\be\label{eqn:prad}
\delta p_{\rm rad}\sim \rho_0 \frac{N_0\om\xi_z}{k_\perp}.
\ee
We assume the convective eddy can only effectively couple to an IGW if the frequencies and horizontal wavelengths match, which requires $\ome\sim\om$ and $k_\perp\sim h^{-1}$.

A large number of convective eddies  contribute to driving a given standing IGW.   This is particularly true for $k_\perp H \gg 1$ and/or $\omega \gg \omega_c$ because then small eddies with sizes $h \ll H$ are responsible for the driving.    The number of eddies contributing to the excitation of a given standing wave is 
\be 
\mathcal{N} \sim \frac{\A dz}{h^3} \sim (\A H^{-2}) \ \left(k_\perp H\right)^{-1} \left(\frac{\om}{\omc}\right)^{9/2}
\label{eqn:Neddy}
\ee
where we have assumed $\omega \gtrsim \omega_c$ and that the excitation happens in a region with thickness $dz \sim k_\perp^{-1}$ (see also eqn.~\ref{eqn:Egend}).  Because an individual IGW is excited by many uncorrelated eddies, the effective pressure fluctuation driving a wave is reduced by a factor of $\sqrt{\mathcal{N}}$ relative to that given in eqn.~\ref{eqn:pconv}.

When $N^2$ is discontinuous at $z=0$ one of the boundary conditions is that $\delta p$ is continuous at $z=0$, so that $\delta p_{\rm rad}\sim \delta p_{\rm conv}$.  Using eqns.~\ref{eqn:pconv}-\ref{eqn:Neddy}, we find that the amplitude of a mode with frequency $\sim\om$ and wavenumber $\sim k_\perp$ is
\be
\xi_z^D\sim \frac{\delta p_{\rm conv} k_\perp}{\rho_0 N_0\om \sqrt{\N}}\sim H^2 k_\perp \frac{\omc^3}{N_0\om^2} \ \N^{-1/2}.
\ee
However, there are $\A k_\perp^2$ such modes in the domain (we have already implicitly summed over the vertical modes in deriving eqns.~\ref{eqn:pconv}, \ref{eqn:prad}), so the typical rms vertical displacement is
\be
\xi_z^D\sim H^2 k_\perp \frac{\omc^3}{N_0\om^2} \ \sqrt{\frac{\A k_\perp^2}{\N}}\sim H \frac{\om}{N_0} (k_\perp H)^{5/2}\left(\frac{\om}{\omc}\right)^{-21/4},
\ee
the same result as in the inhomogeneous wave equation calculation (eqn.~\ref{eqn:dxi}).

\section{Comparison with Previous Work}\label{sec:pastresults}

In this section, we discuss the relationship between our results and previous calculations in the literature.  We begin with GK90, who only consider the discontinuous $N^2$ case.  GK90 solved the fully compressible inhomogeneous wave equation by expanding the perturbation in terms of normal modes and then deriving an amplitude equation.  This is equivalent to our Green's function method (see Appendix~\ref{sec:norm}).  Their end result is very similar to our own; for $k_\perp H\ll 1$ they find \be \frac{d F}{d\log\om \ d\log k_\perp} \sim \M \, \rho_0 u_c^3 (k_\perp H)^3 \left(\frac{\om}{\omc}\right)^{-13/2} \, \mbox{(GK90; eq.~73)} \ee This differs from our result (eqn.~\ref{eqn:dflux}) by a factor of $k_\perp H$.\footnote{Although our final results are similar, there are some ambiguities in GK90's derivation.  In deriving their eqn.~48 from their eqn.~45, GK90 appear to assume that the $\delta p$ are orthogonal under the weighting function $c^{-2}$ and that $\int dz\rho_0 c^{-2}|\delta p|^2\sim 1$.  Both of these are true for sound waves, the main focus of their paper.  However, for IGWs, $\int dz \rho_0 c^{-2}|\delta p|^2\sim \M^2$, and the $\delta p$ are only orthogonal under the weighting function $1$ (see Appendix~\ref{sec:norm} for further discussion on orthogonality).}  We arrive at a different IGW flux because in the Boussinesq approximation $\partial_z\xi_z\sim k_\perp \xi_z$, whereas for the fully compressible system, $\partial_z\xi_z\sim \xi_z/H$ when $k_\perp H\ll 1$.  Accounting for both $k_\perp H\gtrsim 1$ and $k_\perp H \lesssim 1$, the correct scaling of the IGW flux with $k_\perp H$ is $F\sim (k_\perp H)^3 (1+k_\perp H)$.  This does not influence the flux of IGWs which do not break in our smooth $N^2$ calculations.

Because GK90 solve the fully compressible equations, they include multiple scale heights in their convection zone.  They find that the most efficient excitation of waves with frequency $\om$ is at the height where the turnover frequency of the energy bearing eddies is about equal to $\om$.  This effect would be straightforward to include in our model---one would need to derive a Green's function based on the fully compressible eigenfunctions, and then convolve with a vertically varying source term.

GLS91 use a pressure balance argument to study the discontinuous $N^2$ case.  Their power spectrum agrees with eqn.~\ref{eqn:dflux} when $\om\sim \omc$ and $k_\perp H\sim 1$, but not at higher frequencies or wave numbers.  They assume that the pressure perturbation in the convection zone equals the pressure perturbation in the radiative zone.  They take $\delta p_{\rm conv} \sim \rho_0 u_h^2$, and $\delta p_{\rm rad}\sim \rho_0\left(\om \xi_\perp\right)^2$.  This expression for the pressure perturbation in the radiative zone does not satisfy the polarization condition $\delta p_{\rm rad}\sim \rho_0(N_0 \om/k_\perp) \xi_z$ (eqn.~\ref{eqn:raddp}), unless $\xi_\perp \sim k_\perp^{-1}$.  Because many eddies contribute to the excitation of a single IGW mode, GLS91 also decrease their IGW amplitude by a factor of $1/\sqrt{\N}$.  However, they only account for the incoherent sum of small eddies at the interface producing perturbations on large spatial scales.  This gives $\N_{\rm GLS91}\sim (k_\perp h)^{-2}$, where $k_\perp h \ll 1$.  In our analysis, we include eddies which are a distance $k_\perp^{-1}$ above the interface, and we take into account that IGWs excited in different parts of the domain incoherently interfere with each other as they propagate in the radiative zone.  These additional effects yield $\N\sim \A k_\perp^{-1}/h^3$.

P81 uses two different techniques to calculate the IGW flux.  The first uses a pressure balance argument.  Press uses that $\delta p_{\rm conv}\sim\rho_0 u_h^2$, and that $\delta p_{\rm rad}\sim (\rho_0 N_0\om/k_\perp)\xi_z$, and that these pressure perturbations are about equal at the interface.  Throughout his analysis, Press assumes $k_\perp^{-1}\sim h$.  Thus, Press finds
\be\label{eqn:press}
\xi_z \sim \frac{\left(k_\perp h\right)^2}{k_z}\sim \frac{1}{k_z}, \ \ \ \ \ \mbox{(P81; eq.~75)}
\ee
This is the same result given by GLS91, and is consistent with our own assuming $k_\perp h\sim 1$.  This is because $\A k_\perp^2/\N\sim 1$ when $k_\perp h\sim 1$.

Press also derives this result more rigorously using the method of variation of parameters, which is equivalent to using a Green's function.  In addition, Press considers the case in which $N^2$ is continuous at the interface.  He only treats this case in the limit in which $\om\sim N_0$, and finds
\be
\xi_z\sim \frac{1}{k}, \ \ \ \ \ \mbox{(P81; eq.~88)}
\ee
the same result as eqn.~\ref{eqn:press}.  However, note that if $\om\sim N_0$, then $N_0 k_\perp d/\om\ll 1$, and the smooth result cannot be used (these waves see the interface as discontinuous).  In addition, Press's use of standard WKB matching to treat the smooth $N^2$ profile is generally not applicable (see Appendix~\ref{sec:tanh}).

Finally, we consider the work of B09.  In their paper, Belkacem et al. numerically calculate the eigenfunctions for a solar structure model, use a convection simulation to specify the source term, and solve an amplitude equation in the same way as GK90.  It is unclear whether the $N^2$ profile in their solar structure model has a smooth transition between the radiative and convection zones---if their $N^2$ profile is discontinuous (Section~\ref{sec:discontinuous}) or has an abrupt transition (Appendix~\ref{sec:linear}), they will derive different eigenfunctions than for a $\tanh$ profile (Appendix~\ref{sec:tanh}).  These eigenfunctions will produce a smaller flux (eqns.~\ref{eqn:dfluxt}, \ref{eqn:lfluxr}) than we predict for a very smooth radiative-convective transition (eqn.~\ref{eqn:sfluxr}).

Another key difference is that Belkacem et al. use an eddy-time correlation function $\chi_{\vec{k}}(\om)$, which in the notation of this paper can be written as $\chi(\om;\ome)$.  This function describes how efficiently an eddy with size $1/k$ and turn-over frequency $\ome=u_{\vec{k}} k$ excites a wave with frequency $\om$.  Our analysis implicitly assumes $\chi(\om;\ome)\sim \exp(-\om^2/\ome^2)$.  This Gaussian eddy-time correlation function implies that eddies with turn-over frequencies $\ome$ only excite waves with frequencies $\om$.  However, the turbulence in the convection simulation in B09 is not well described by a Gaussian eddy-time correlation function.  Instead, Belkacem et al. find that a Lorentzian distribution, $\chi(\om;\ome)\sim (1+2(\om/\ome)^2)^{-1}$, is more accurate.  This indicates that waves with frequency $\om$ can be excited by a broad range of eddies.  In general, this makes wave excitation more efficient.  It would be straightforward to generalize our results to this Lorentzian expression for $\chi(\om;\ome)$.

\section{Discussion \& Conclusions}\label{sec:conclusion}

In this paper we have calculated the excitation of internal gravity waves (IGW) by turbulent convection, motivated by the application to stellar convection.  We assume that the source term exciting the IGWs can be modeled by Reynolds stresses associated with uncorrelated eddies in a Kolmogorov turbulent cascade.  Our main results are the IGW fluxes, eqns.~\ref{eqn:dfluxt}, \ref{eqn:sfluxr} \& \ref{eqn:lfluxr}.  In particular, we predict a {\it larger wave flux than previous calculations for low frequency waves} which satisfy $N_0 k_\perp d/\om\gg 1$, where $N_0$ is the buoyancy frequency in the radiative zone, $k_\perp$ and $\om$ are the horizontal wavenumber and frequency of the IGW, respectively, and $d$ is the thickness of the transition region between the radiative and convection zones.  We also reconcile somewhat disparate claims in the literature by showing that different methods, such as pressure balance arguments and solving the inhomogeneous wave equation, predict the same IGW power spectrum when using the same assumptions (Section~\ref{sec:pressure}).

An IGW with frequency $\om$ sees the transition between the radiative and convection zones as discontinuous if $N_0 k_\perp d/\om \ll 1$.  In this case, the total flux is $F^D\sim F_{\rm conv} \ \M$ (eqn.~\ref{eqn:dfluxt}), as derived in past work, where $F_{\rm conv}$ is the convective flux and $\M$ is the convective Mach number.  The most efficiently excited waves have frequencies $\om\sim\omc$, the eddy turn-over frequency of the largest turbulent eddies, and $k_\perp \sim H^{-1}$, the inverse of the pressure scale height.  These most efficiently excited waves are marginally susceptible to wave breaking when they enter the radiative region.

If, however, the transition between radiative and convective regions is smooth (i.e., $N_0 k_\perp d/\om\gg 1$), the problem becomes more complicated.  The IGW flux depends on the structure of the buoyancy frequency $N^2(z)$ near the transition between the radiative and convective regions.  We parameterize the transition using both a $\tanh$ profile, which is we believe represents the smoothest possible transition, and a piecewise linear profile, which is the most abrupt transition possible.  These two examples bound the physical possibilities, and we expect real $N^2$ profiles in stars to be somewhere in between.  The wave excitation is more efficient when $N^2$ is smooth because the IGW eigenfunctions change amplitude rapidly near the interface (as originally discussed by P81).

The total IGW fluxes for the $\tanh$ and piecewise linear profiles are $F^T\sim F_{\rm conv} (d/H)\gg F^D$ (eqn.~\ref{eqn:fluxtt}), and $F^L\sim F_{\rm conv} \M^{2/3} (d/H)^{1/3}\gg F^D$ (eqn.~\ref{eqn:dfluxl}), respectively.  Again, the most efficiently excited waves have frequencies $\om\sim\omc$ and $k_\perp\sim H^{-1}$.  However, these waves are extremely prone to wave breaking, as $k_z\xi_z\gg 1$ in the radiative region (e.g., P81).  These waves will break in the transition region between the radiative and convection zones.  The flux of IGWs that are marginally susceptible to wave breaking (i.e., have $k_z\xi_z\sim 1$) is $F^T\sim F_{\rm conv} \M^{5/8} (d/H)^{3/8}$ (eqn.~\ref{eqn:sfluxr}) and $F^L\sim F_{\rm conv} \M^{57/65} (d/H)^{8/65}$ (eqn.~\ref{eqn:lfluxr}).  This is larger than the discontinuous $N^2$ flux by $(\M H/d)^{-3/8}$ for the $\tanh$ profile, and by $(\M H/d)^{-8/65}$ for the piecewise linear profile.

In the Sun, $\omc\sim10^{-3} N_0$, so $\M\sim10^{-3}$ \citep[e.g.,][]{BVZ12}, and $d$ is estimated to be $\sim 0.1 H$ \citep{CD11}.  IGWs produced by the energy bearing eddies have $N_0 k_\perp d/\om\sim 10^2$, and thus the transition region must be treated as smooth.  This suggests that the IGW flux in the Sun is somewhere between
\bea
&& F^T\sim F_{{\rm conv}}\M^{5/8} \left(\frac{d}{H}\right)^{3/8}\sim 5\times 10^{-3} \ F_{{\rm conv}}, \\
&& F^L\sim F_{\rm conv} \M^{57/65} \left(\frac{d}{H}\right)^{8/65}\sim 2\times 10^{-3} \ F_{\rm conv},
\eea
about two to five times larger than the flux in the discontinuous $N^2$ case.  In both cases, the flux is dominated by waves with frequencies near $\omc$, and wave numbers near $H^{-1}$.

 We expect the $N^2$ profile in stars to be somewhere between the $\tanh$ profile and the piecewise linear profile.  Real $N^2$ profiles are likely to have continuous derivatives, which precludes the piecewise linear profile.  However, a piecewise linear function can be smoothed over an arbitrarily small length scale to form an infinitely differentiable function.  Indeed, in simulations of penetrative convection, the time and spatially averaged $N^2$ profile appears similar to a $\tanh$ profile (e.g., Fig.~3 in \citealt{RG06} and Fig.~7 in \citealt{RG052}).  Specifically, these simulations find that $dN^2/dz|_{z_i}\ll N_0^2/d$, i.e., the slope of $N^2$ near $N^2=0$ is much less than in a simple piecewise linear model.  This suggests that even if real $N^2$ profiles look closer to piecewise linear, the appropriate value for $d$ might be much larger than expected.  For these reasons, we expect IGW generation in stars to more closely follow the $\tanh$ profile results than the piecewise linear results.

In this paper we have also briefly considered IGW excitation due to turbulence driven by overshooting convective plumes (Section~\ref{sec:overshoot}).  These results depend sensitively on our assumptions regarding the turbulence within the overshoot region, which is poorly understood.  However, our calculations suggest that IGW excitation is about as efficient in the overshoot region as in the convection zone.  The flux in the energy-bearing mode, using solar parameters, is smaller in the overshoot region by a factor of $0.2$, but the flux in some higher frequency modes can be slightly larger in the overshoot region.  These higher frequency IGWs are the ones most likely to be observed in main sequence stars \citep[e.g.][]{2012arXiv1210.5525S}, making it important to understand excitation in the overshoot region in more detail in future work.  Modes which have $k_\perp d\gg 1$ are excited much more efficiently in the overshoot region than in the convection zone, where they are exponentially suppressed.  It is difficult to excite the large, energy-bearing modes in the overshoot region, because $k_z$ is larger in the overshoot region than in the convection zone.  Thus, only smaller eddies can couple to the large modes, decreasing the IGW flux produced in the overshoot region.

The increase in wave flux due to a smooth radiative-convective interface is only for waves with $N_0 k_\perp d/\om\gg 1$, i.e., for low frequency waves.  For certain applications (e.g., helioseismology), the flux of low frequency waves is unimportant.  In particular, low frequency $g$-modes in the Sun and massive stars are strongly damped by radiative diffusion and are unlikely to be seen at the surface.  Thus, the increase in wave flux we predict for low frequency waves does not change the expected amplitudes of potentially observable $g$-modes in main sequence stars.

However, low frequency waves are important for the angular momentum transport, mixing, and/or mass loss due to IGWs excited by stellar convection.  For example, a larger IGW flux may increase the predicted mass loss in the final stages of the life of a massive star \citep{QS12} and in Type Ia supernova progenitors \citep{Piro11}.  This will be studied in detail in future work.

We have shown that there is significant wave breaking near the radiative-convective interface if $N^2$ is smooth.  Wave breaking produces turbulence and can lead to additional IGW generation (Fritts 2009).  When $N^2$ is smooth, the flux in modes which are unstable to breaking is a significant fraction of $F_{\rm conv}$; thus the breaking process has the potential to excite a non-negligible flux of IGWs.  In addition, wave breaking could redistribute energy in $(k_\perp,\om)$ space, thus potentially modifying the IGW power spectrum from that calculated here.

In order to make a more accurate prediction of the wave flux and spectrum, one would need to use a stellar structure model with a realistic radiative-convective interface and a better representation of the convective turbulence, as in B09.  Our results highlight the importance of adequately resolving the smooth transition between the radiative and convective regions in such calculations.  A discontinuous or abrupt transition will give a different IGW flux than a smooth transition. We note that the radiative-convective transition seen in numerical simulations of penetrative convection is significantly smoother than the transition in typical 1D stellar models \citep[e.g.,][]{RG06}.

Perhaps the most promising way to test the results of this paper is through comparison with direct numerical simulations of a radiative zone adjacent to a convection zone \citep[e.g.,][]{RG05, BMT11}.  Although such simulations typically require artificially high conduction in the radiative zone, and it is unclear how to best identify IGWs \citep{DBN05}, this is probably the simplest system in which one can quantify the IGW flux generated by convection.  We hope that analysis of such simulations can provide a quantitative test of the theory derived in this paper in the near future.

\acknowledgments
This material is based upon work supported by the National Science Foundation Graduate Research Fellowship under Grant No. DGE 1106400.  DL acknowledges support from a Hertz Foundation Fellowship.  This work was also partially supported by NASA HTP grant NNX11AJ37G, a Simons Investigator award from the Simons Foundation to EQ, the David and Lucile Packard Foundation, and the Thomas Alison Schneider Chair in Physics at UC Berkeley.  We especially thank G. Vasil for helping solve for the eigenfunctions in Appendix~\ref{sec:tanh} and A. Lieb for helping understand the differences between the mode projection and Green's function formalisms.

\appendix

\section{tanh Profile Eigenfunctions}\label{sec:tanh}

We will derive the eigenfunctions for the equation
\be\label{eqn:ode}
\frac{\partial^2}{\partial z^2}\xi_z +\left(\frac{N^2(z)}{\om^2}-1\right)k_\perp^2\xi_z=0,
\ee
where
\be\label{eqn:tanhN2}
N^2(z)= \frac{N_0^2+\omc^2}{2}\left(\tanh\left(-\frac{z}{d}\right)+1\right) -\omc^2.
\ee
The transition between oscillatory behavior and exponential behavior (where $N^2(z)=\om^2$) is at $z_t$ given by
\be
\frac{\om^2+\omc^2}{N_0^2+\omc^2}\sim \exp\left(-2 \frac{z_t}{d}\right).
\ee
The eigenfunction in the radiative zone is well approximated by the WKB solution,
\bea
\xi_z= B_1\left(N_0k_\perp/\om\right)^{1/2} \ k_z(z)^{-1/2} \cos\left( \int dz k_z(z) +\pi/4\right) \nonumber \\
+B_2\left(N_0k_\perp/\om\right)^{1/2} \ k_z(z)^{-1/2} \sin\left( \int dz k_z(z) +\pi/4\right), \label{eqn:wkbsoln}
\eea
where we define the vertical wavenumber to be
\be
k_z^2(z)=k_\perp^2\left(N^2(z)/\om^2-1\right).
\ee
Near $z_t$, the WKB solution in the radiative region diverges.  We wish to derive a new set of functions which closely approximate the eigenfunctions for $z>z_t$.

In many problems, the WKB solutions near a turning point can be asymptotically matched onto Airy functions, which provide a connection between exponentially decaying and oscillatory WKB solutions.  However, we cannot use this approach when $k_\perp d < 1$; in this parameter regime $k^2_z(z)$ cannot be well approximated as linear near $z_t$.  Instead, we will show that when $k_\perp d<1$ the eigenfunctions can be well approximated in terms of Bessel functions.  Furthermore, these Bessel function solutions are also a good approximation when $k_\perp d \geq 1$.

To show this, first note that if $\exp(-2z/d)\ll 1$, we can approximate
\be\label{eqn:N2approx}
N^2(z) \approx \left(N_0^2+\omc^2\right) \exp\left(-\frac{2z}{d}\right)-\omc^2.
\ee
The solutions to the wave equation (eqn.~\ref{eqn:ode}) for this approximate $N^2(z)$ function are
\bea
\xi_z = C_1 J_{ \bar{\om}k_\perp d/\om}\left(k_\perp d \frac{\sqrt{N_0^2+\omc^2}}{\om} \exp\left(-\frac{z}{d}\right)\right) \nonumber \\
 + C_2 Y_{\bar{\om}k_\perp d/\om}\left( k_\perp d \frac{\sqrt{N_0^2+\omc^2}}{\om} \exp\left(-\frac{z}{d}\right)\right), \label{eqn:bessel}
\eea
where $J$ and $Y$ are the Bessel functions of the first and second kind, respectively.  We have also defined $\bar{\om}^2=\om^2+\omc^2$, where $\bar{\om}/\om$ ranges between $\sqrt{2}$ and $1$.  These approximate the solution for large positive $z$.  We can asymptotically match the Bessel functions onto the WKB solution in the radiative zone (eqn.~\ref{eqn:wkbsoln}).  We will make use of the following asymptotic forms for $J$ and $Y$:
\bea
J_\alpha(x) \sim \frac{1}{\Gamma (\alpha+1)}\left(\frac{x}{2}\right)^\alpha, \\
Y_\alpha(x) \sim -\frac{\Gamma(\alpha)}{\pi} \left(\frac{2}{x}\right)^\alpha,
\eea
provided that $0<x\ll\sqrt{\alpha+1}$, and
\bea
J_\alpha(x) \sim \sqrt{\frac{2}{\pi x}} \cos\left( x-\frac{\alpha\pi}{2}-\frac{\pi}{4}\right), \\
Y_\alpha(x)\sim \sqrt{\frac{2}{\pi x}} \sin\left( x -\frac{\alpha\pi}{2}-\frac{\pi}{4}\right),
\eea
provided that $x\gg |\alpha^2+1/4|$.

We must consider two regimes, depending on the size of $k_\perp d$.  First consider $k_\perp d \ll 1$.  We can use the asymptotic formula for large arguments provided that
\be
k_\perp d \frac{\sqrt{N_0^2+\omc^2}}{\om}\exp\left(-\frac{z}{d}\right) \gg \frac{1}{4}.
\ee
This constraint can be satisfied simultaneously with $\exp(-2z/d)\ll 1$, implying that the asymptotic form of the Bessel functions are good approximations to the eigenfunctions.  If we approximate $k_z^2(z)$ as
\be
k_z^2(z)\approx k_\perp^2 \frac{N_0^2+\omc^2}{\om^2}\exp\left(-\frac{2z}{d}\right),
\ee
we can approximate eqn.~\ref{eqn:bessel} by
\bea
\xi_z\approx C_1 \sqrt{\frac{2}{\pi d}} \left(k_z^2(z)\right)^{-1/4} \cos\left( - d \left( k_z^2(z)\right)^{1/2} + \frac{\pi \bar{\om} k_\perp d }{2\om} +\frac{\pi}{4}\right) \nonumber \\
-C_2\sqrt{\frac{2}{\pi d}} \left(k_z^2(z)\right)^{-1/4} \sin\left( - d \left( k_z^2(z)\right)^{1/2} + \frac{\pi \bar{\om} k_\perp d }{2\om} +\frac{\pi}{4}\right).
\eea
This matches onto the WKB solution in the radiative region since $k_\perp d$ is small.  The amplitudes are
\bea\label{eqn:kpd<1,1}
&& C_1=B_1 \left(\frac{\pi N_0 k_\perp d}{2\om}\right)^{1/2}, \\
&& C_2=-B_2 \left(\frac{\pi N_0 k_\perp d}{2 \om}\right)^{1/2}. \label{eqn:kpd<1,2}
\eea

Now assume $k_\perp d \gg 1$.  In this case, the asymptotic form of the Bessel functions for small argument is only valid when $\exp(-z/d)\gg 1$, i.e., for positions where the Bessel functions themselves are not a good approximation to the eigenfunctions ($N^2(z)$ cannot be simplified as in eqn.~\ref{eqn:N2approx} if $\exp(-z/d)\gg 1$).  However, in this limit we can use the WKB approximation in the convective region, and connect the two WKB solutions with Airy functions.  Thus, in the convective region, we have
\bea
\xi_z\sim \left(B_1/2\right) (N_0 k_\perp/\om)^{1/2} k_z(z)^{-1/2} \exp \left(-\int_{z_t}^z dz' |k_z(z')|\right) \nonumber \\
+ B_2 (N_0 k_\perp/\om)^{1/2} k_z(z)^{-1/2} \exp\left(+\int_{z_t}^z dz' |k_z(z')|\right).
\eea
For $z$ much larger than $z_t$, this becomes
\bea
\xi_z \sim \frac{B_1}{2} \left(\frac{N_0}{\om}\right)^{1/2} \left(\frac{e}{2}\right)^{\bar{\om}k_\perp d/\om} \exp(-(z-z_t)k_\perp \bar{\om}/\om) \nonumber \\
+ B_2 \left(\frac{N_0}{\om}\right)^{1/2} \left(\frac{2}{e}\right)^{\bar{\om}k_\perp d/\om} \exp(+(z-z_t)k_\perp\bar{\om}/\om),
\eea

For $z$ much larger than $z_t$, the Bessel functions are a good approximation to the eigenfunction.  In the limit of large $z$, the Bessel functions become
\bea
\xi_z\sim  C_1 \left(\frac{1}{e\pi k_\perp d}\right)^{1/2} \left(\frac{e\bar{\om}}{2\om}\right)^{\bar{\om}k_\perp d/\om+1/2} \exp\left(-\frac{(z-z_t)k_\perp\bar{\om}}{\om}\right) \nonumber \\
-C_2 \left(\frac{4}{e \pi k_\perp d}\right)^{1/2}  \left(\frac{2 \om}{e \bar{\om}}\right)^{\bar{\om}k_\perp d/\om+1/2} \exp\left(+\frac{(z-z_t)k_\perp\bar{\om}}{\om}\right).
\eea
Thus, the Bessel function solution matches onto the WKB solution in the convective region when
\bea\label{eqn:kpd>1,1}
C_1=B_1\left(\frac{\pi N_0 k_\perp d}{2\om}\right)^{1/2} \left(\frac{\om}{\bar{\om}}\right)^{\bar{\om}k_\perp d/\om+1/2}, \\
C_2=-B_2\left(\frac{\pi N_0 k_\perp d}{2\om}\right)^{1/2} \left(\frac{\bar{\om}}{\om}\right)^{\bar{\om}k_\perp d/\om+1/2}. \label{eqn:kpd>1,2}
\eea

Using eqns.~\ref{eqn:kpd<1,1}, \ref{eqn:kpd<1,2} \& \ref{eqn:kpd>1,1}, \ref{eqn:kpd>1,2}, we can approximate $\xi_z$ by
\bea
\xi_z\sim B_1 \left(\frac{\pi N_0k_\perp d}{2\om}\right)^{1/2} \left(\frac{\om}{\bar{\om}}\right)^{\bar{d}} J_{\bar{d}}\left(k_\perp d\frac{\sqrt{N_0^2+\omc^2}}{\om}\exp\left(-\frac{z}{d}\right)\right) \nonumber \\
+B_2 \left(\frac{\pi N_0k_\perp d}{2\om}\right)^{1/2} \left(\frac{\bar{\om}}{\om}\right)^{\bar{d}} Y_{\bar{d}}\left(k_\perp d\frac{\sqrt{N_0^2+\omc^2}}{\om}\exp\left(-\frac{z}{d}\right)\right),
\eea
where we have defined $\bar{d}=\bar{\om}k_\perp d/\om$.  This will be a good approximation for $\xi_z(z)$ as long as $\exp(-z/d)\ll 1$.

For the purposes of determining the convective excitation of IGWs, we are interested in evaluating $\xi_z$ between $z_i$ and $z_i+1/k_\perp$, where $z_i$ is the location of the interface between the radiative and convective regions.  Since
\be
\frac{\omc^2}{N_0^2+\omc^2}\sim\exp\left(-2\frac{z_i}{d}\right),
\ee
the argument of the Bessel functions varies from $\omc k_\perp d/\om$ to $\exp(-(k_\perp d)^{-1} ) \omc k_\perp d/\om$.  Within this range, the Bessel functions change by about a factor of $e$.  It is thus within the accuracy of our calculation to take $\xi_z$ to be about constant within this range:  at $z=z_i$, we have that
\bea
\xi_z \sim B_1 \left(\frac{\pi N_0k_\perp d}{2\om}\right)^{1/2} \left(\frac{\om}{\bar{\om}}\right)^{\bar{\om}k_\perp d/\om} J_{\bar{\om}k_\perp d/\om}\left(\omc k_\perp d/\om\right) \nonumber \\
+B_2 \left(\frac{\pi N_0k_\perp d}{2\om}\right)^{1/2} \left(\frac{\bar{\om}}{\om}\right)^{\bar{\om}k_\perp d/\om} Y_{\bar{\om}k_\perp d/\om}\left(\omc k_\perp d/\om\right). \label{eqn:tanhsolution}
\eea
In evaluating eqn.~\ref{eqn:tanhsolution}, we need to calculate $J_x(xa)$, where $x=\bar{\om} k_\perp d/\om$, and $a=\omc/\bar{\om}<1/\sqrt{2}$.  A good set of approximations for the Bessel functions for $x\ll 1$ and $x\gg 1$ is given in eqn.~\ref{eqn:besselapprox} of the main text (based on expansions of $J_x(xa)$ from \citet{AS72}).

\subsection{Numerical Verification}\label{sec:numeric}

Here we will present numerical verification of our approximate solutions in the above subsection.  We numerically integrated the homogeneous differential equation (eqn.~\ref{eqn:ode}) with $N^2$ given by eqn.~\ref{eqn:tanhN2} in Mathematica using the ``ImplicitRungeKutta'' method, and solved for a physical solution, satisfying $\xi_z\rightarrow 0$ as $z\rightarrow \infty$ (see Fig.~\ref{fig:eigenfunction} in the main text for a representative eigenfunction).  We pick the right boundary to be a point $b$ deep within the convective region, where $k_z^2(b)=-k_\perp^2$, specify $\xi_z(b)=1$, $\xi_z'(b)=-k_\perp$, and integrate $\xi_z$ leftwards into the radiative region.  This ensures that $\xi_z$ satisfies the boundary condition $z\rightarrow+\infty$.  We find that our calculations are insensitive to the value of $b$, provided that it is sufficiently larger than $z_t$.

To test the approximations described in the above subsection, we calculate the value of the physical eigenfunction at the interface between the radiative and convective regions $\xi_z(z_i)$.  Because any multiple of the eigenfunction is also an eigenfunction, we normalize by $B_1$ (see eqn.~\ref{eqn:wkbsoln}), which is the amplitude of the oscillations deep in the radiative zone.  Equation~\ref{eqn:tanhsolution} predicts
\be
\xi_z(z_i)/B_1=\left(\frac{\pi N_0k_\perp d}{2\om}\right)^{1/2} \left(\frac{\om}{\bar{\om}}\right)^{\bar{\om}k_\perp d/\om}J_{\bar{\om}k_\perp d/\om}\left(k_\perp d\frac{\omc}{\om}\right).
\ee
Our analysis is only valid if we are in the smooth $N^2$ limit, i.e., if $N_0k_\perp d/\om\gg 1$.

\begin{figure}
  \begin{center}
    \includegraphics[width=\linewidth]{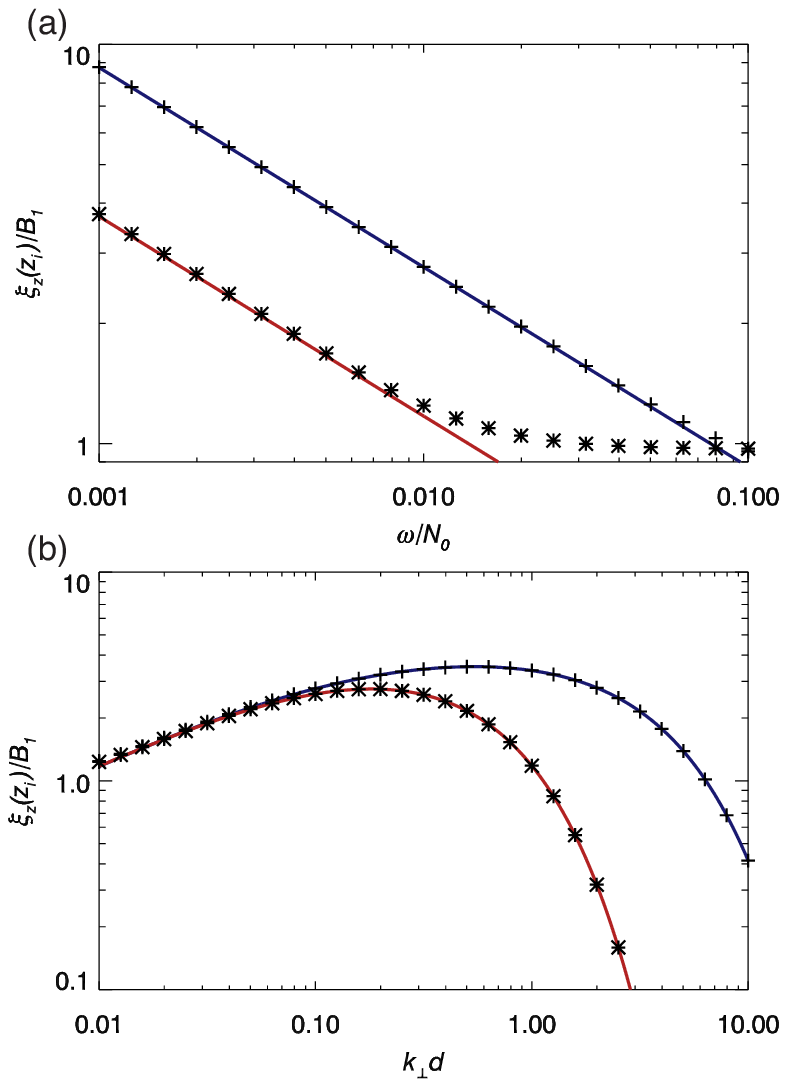}
  \end{center}
\caption{
The normalized eigenfunction at the radiative-convective interface $z_i$.  The symbols denote the numerical solution, and the lines denote the analytic prediction, eqn.~\ref{eqn:tanhsolution}.  In the top panel, we vary $\om/N_0$, fixing $\omc=\om$.  The blue line and crosses have $k_\perp d=0.1$, and the red line and asterisks have $k_\perp d=0.01$.  The numerical solution matches the analytic prediction for smooth $N^2$ when $N_0 k_\perp d/\om\gg 1$, and approaches one (the discontinuous $N^2$ solution) when $N_0 k_\perp d/\om\ll 1$.  In the bottom panel, we vary $k_\perp d$, fixing $\om/N_0=0.01$ and setting $\omc/N_0=0.01$ (blue curve, crosses) or $\omc/N_0=0.002$ (red curve, asterisks).  Again, there is good agreement between the numerical solution and the analytic prediction.
  \label{fig:amp}
}
\end{figure}

In Figure~\ref{fig:amp} we compare our numerical results to the analytic predictions.  In Figure~\ref{fig:amp} (top panel) we vary $\om/N_0$ for two different values of $k_\perp d$.  The numerical solutions agree with our prediction when $N_0 k_\perp d/\om\gg 1$.  In the opposite limit, when $N_0 k_\perp d/\om\ll 1$, we can treat $N^2$ as discontinuous, so $\xi_z$ is continuous across the interface, and $\xi_z(z_i)/B_1=1$, as is the case for the lower curve in Figure~\ref{fig:amp} (top panel).  In Figure~\ref{fig:amp} (bottom panel) we vary $k_\perp d$, fixing $\om/N_0=0.01$, for two values of $\omc/N_0$.  In this case, we have $N_0 k_\perp d/\om=1$ when $k_\perp d=0.01$.  The normalized eigenfunctions approach one as $k_\perp d$ decreases, and the numerical solutions begin to deviate slightly from the analytic prediction near $k_\perp d =0.01$.  These results indicate that our analytic solution for $\xi_z$ near $z_i$ is accurate provided we are in the smooth $N^2$ limit.  The numerical solutions also show how the eigenfunctions transition between the smooth and discontinuous $N^2$ limits.

\section{Piecewise Linear $N^2$}\label{sec:linear}

In the limit of smooth $N^2$, the eigenfunctions, Green's function, and IGW flux all depend on the nature of the transition between radiative and convective regions.  In this paper, we focus on the case of a $\tanh$ profile (Appendix~\ref{sec:tanh}), as we think it is the best simple model of this transition region.  However, in this appendix, we consider another analytically tractable transition---a piecewise linear $N^2$ profile.  This is the most abrupt transition possible, and thus provides a lower limit to the efficiency of wave excitation for a ``smooth'' radiative-convective transition.

We assume $N^2$ is given by
\be
N^2(z)=\left\{
	\begin{array}{ll}
		N_0^2  & \mbox{if } z \leq -d/2, \\
		(N_0^2-\omc^2)/2 - (N_0^2+\omc^2) \ (z/d) & \mbox{if } -d/2<z<d/2, \\
		-\omc^2 & \mbox{if } z \geq d/2.
	\end{array}
\right.
\ee
We have that $N^2(z)=\om^2$ at the point
\be
z_t= \frac{N_0^2-2\om^2-\omc^2}{N_0^2+\omc^2} \ \left(\frac{d}{2}\right),
\ee
and that $N^2(z)=0$ at
\be
z_i=\frac{N_0^2-\omc^2}{N_0^2+\omc^2} \ \left(\frac{d}{2}\right).
\ee

The solutions in each region are
\bea\label{eqn:linearrad}
\xi_z & = & B_1\cos(N_0k_\perp (z+d/2)/\om) + B_2\sin(N_0k_\perp (z+d/2)/\om), \\
& & \ \ \ \ \ \mbox{for } z<-d/2, \nonumber \\ \label{eqn:lineartrans}
\xi_z & = & C_1\exp(-k_\perp (z-d/2)) + C_2\exp(k_\perp (z-d/2)), \\
& & \ \ \ \ \ \mbox{for } z>d/2, \nonumber \\ \label{eqn:linearconv}
\xi_z & = & D_1\Ai\left(K_1^{1/3} (z-z_t)\right) + D_2\Bi\left(K_1^{1/3}(z-z_t) \right), \\
& & \ \ \ \ \ \mbox{for } -d/2<z<d/2, \nonumber
\eea
where $\Ai$, $\Bi$ are the Airy functions of the first and second kind, and
\be
K_1=\left.\frac{dk_z^2(z)}{dz}\right|_{z_t}=\frac{k_\perp^2}{d}\frac{N_0^2+\omc^2}{\om^2}.
\ee
We can relate the six coefficients in eqns.~\ref{eqn:linearrad}-\ref{eqn:linearconv} to one another using four boundary conditions:  $\xi_z$ and $\xi_z'$ must be continuous at $z=\pm d/2$.

First consider the boundary at $z=+d/2$.  The argument of the Airy functions at this boundary is
\be
\left(k_\perp d\right)^{2/3}\left(\frac{\om^2+\omc^2}{\left(N_0^2+\omc^2\right)^{2/3} \left(\om^2\right)^{1/3}}\right)\sim \left(\frac{\om^2k_\perp d}{N_0^2}\right)^{2/3}.
\ee
This is much smaller than one unless $k_\perp d$ is extremely large.  One can check that IGW excitation is exponentially suppressed when $\om^2 k_\perp d/ N_0^2\gg 1$.  Thus, we will assume that $\om^2 k_\perp d/N_0^2\ll 1$.  This implies that $\Ai|_{d/2},\Bi|_{d/2},\Ai'|_{d/2},\Bi'|_{d/2}$ are all of order one, where we have introduced the shorthand $\Ai|_z=\Ai(K_1^{1/3}(z-z_t))$, and similarly for the other functions.  To order of magnitude, we have that
\be
C_1 + C_2 \sim D_1\Ai|_{d/2} + D_2\Bi|_{d/2},
\ee
and
\be
C_1 - C_2 \sim \frac{K_1^{1/3}}{k_\perp} \left(D_1\Ai'|_{d/2}+D_2\Bi'|_{d/2}\right).
\ee
Notice that
\be
K_1^{1/3}/k_\perp \sim \left(\frac{1}{k_\perp d} \frac{N_0^2+\omc^2}{\om^2}\right)^{1/3}\gg 1.
\ee

Now consider the boundary at $z=-d/2$.  The argument of the Airy functions at this boundary is
\be
\left(k_\perp d\right)^{2/3} \left(\frac{N_0^2+\omc^2}{\om^2}\right)^{1/3}\sim \left(\frac{N_0 k_\perp d}{\om}\right)^{2/3}\gg 1,
\ee
where the last inequality follows from assuming that we are in the smooth $N^2$ limit.  We thus have
\bea
&& \xi_z|_{-d/2}\sim \left( \frac{N_0k_\perp d}{\om} \right)^{-1/6} \nonumber \\
&& \times \ \left[ D_1\cos\left( \frac{2}{3} \frac{N_0k_\perp d}{\om} +\frac{\pi}{4}\right) + D_2\sin\left(\frac{2}{3} \frac{N_0k_\perp d}{\om} + \frac{\pi}{4}\right)\right],
\eea
implying
\be
B_1\sim\left(\frac{N_0k_\perp d}{\om}\right)^{-1/6} \left(D_1\cos(\phi)+D_2 \sin(\phi)\right),
\ee
where $\phi=(2/3)(N_0k_\perp d/\om) + \pi/4$.  Similarly, by comparing $\xi_z'$ on either side of $z=-d/2$ we find
\be
B_2\sim\left(\frac{N_0k_\perp d}{\om}\right)^{-1/6} \left(-D_1 \sin(\phi)+D_2\cos(\phi)\right).
\ee

Using these boundary conditions, we find that the physical eigenfunction is
\be\label{eqn:lineig1}
\eta_z^L \sim\left\{
	\begin{array}{ll}
		B_1\cos\left(\frac{N_0 k_\perp (z+d/2)}{\om}\right) + B_2\sin\left(\frac{N_0 k_\perp (z+d/2)}{\om}\right)  &  z < -d/2, \\
		\tilde{B_1}\left(\frac{N_0 k_\perp d}{\om}\right)^{1/6}\exp(-k_\perp(z-d/2)) & z > d/2,
	\end{array}
\right.
\ee
where we use {\it superscript $L$} to denote the eigenfunction for the {\it piecewise linear} $N^2$ profile, and $\tilde{B_1}\sim B_2\sim B_1$.  An unphysical eigenfunction is
\be\label{eqn:lineig2}
\xi_z^L \sim\left\{
	\begin{array}{ll}
		B_2\sin\left(\frac{N_0 k_\perp (z+d/2)}{\om}\right)  &  z < -d/2, \\
		\left(\frac{N_0k_\perp d}{\om}\right)^{-1/6}\frac{N_0}{\om} \ \times \\
		\left(\tilde{B_1}\exp(-k_\perp (z-d/2)) +\tilde{B_2}\exp(k_\perp (z-d/2))\right) & z > d/2,
	\end{array}
\right.
\ee
where $\tilde{B_2}\sim\tilde{B_1}\sim B_2$.  Note that the constants $B_1, B_2$ in $\eta_z^L$ and $\tilde{B_1},\tilde{B_2}$ in $\xi_z^L$ vary sinusoidally with $d$ (as well as the other parameters of the problem).  Thus, although for most values of $d$ they are the same size, there are specific values of $d$ for which one term is much larger than the other.

The Green's function for $z<0$ and $\zeta>0$ is then
\bea
G^L(z,t,\zeta,\tau)\sim\sum_{\om'} \frac{\om'\sqrt{\rho_0}}{N_0 k_\perp^2 \sqrt{L}} \left(\frac{N_0 k_\perp d}{\om'}\right)^{1/6} \nonumber \\
\times \ \xi_z^l(z;\om')\exp(-k_\perp \zeta -i\om' (t-\tau)). \label{eqn:linGF}
\eea

\section{Mode Projection Formalism (GK90)}\label{sec:norm}

In GK90, an amplitude equation is derived by projecting the inhomogeneous wave equation onto specific modes.  We will show that their approach gives the same result as our Green's function approach, provided that the correct inner product is used.

First start with the inhomogeneous equation for $\xi_z$ in the Boussinesq approximation
\be
\grad^2 \frac{\partial^2}{\partial t^2}\xi_z + N^2\grad^2_\perp\xi_z = S.
\ee
In the mode projection formalism, we decompose $\xi_z$ as
\be
\xi_z = \frac{1}{\sqrt{\A}}\sum_{\om'} A(t;\om')\eta_z(z;\om')\exp(ik_xx+ik_yy-i\om't),
\ee
where $\eta_z(z;\om')$ are the physical solutions satisfying the homogeneous wave equation.  Substituting this into the inhomogeneous wave equation, multiplying by $\rho_0\eta_z^*(z;\om)\exp(-ik_xx-ik_yy+i\om t)$ and integrating over $d^3x dt$, we find
\bea
|A(t;\om)|= \frac{\om}{2k_\perp^2\sqrt{\A}} \int_{\ -\infty}^{\ t}d\tau \int dxdy \exp(-ik_xx-ik_yy+i\om\tau)  \nonumber \\
\times \ \int_{\ z_i}^{\ L} d\zeta \rho_0 S(x,y,\zeta,\tau)\eta_z^*(\zeta;\om). \label{eqn:ampxiz}
\eea
A crucial step in deriving this is using
\be
\int dz \rho_0 \partial_z\eta_z(z;\om')\partial_z\eta_z^*(z;\om)=\delta_{\om\om'} \ \frac{k_\perp^2}{\om \om'}.
\ee
That is, the $\eta_z(z;\om)$ are orthogonal with respect to the inner product $\la a,b\ra = \int dz \rho_0 \partial_z a \partial_z b^*$.  This follows from our normalization equation (eqn.~\ref{eqn:normalization}) and the polarization conditions (eqn.~\ref{eqn:radpol}).

Although we use $\xi_z$ as our perturbation variable in this paper, GK90 uses $\delta p$.  The inhomogeneous wave equation for $\delta p$ in the Boussinesq approximation is
\be\label{eqn:deltapode}
\grad^2\frac{\partial^2}{\partial t^2}\delta p + N^2\grad^2_\perp \delta p = \bar{S},
\ee
where $\bar{S}\sim (\rho_0 \om^2/k_\perp) \ S$.  As above, we can decompose $\delta p$ into eigenmodes
\be\label{eqn:deltapdecomp}
\delta p = \frac{1}{\sqrt{\A}}\sum_{\om'} A(t;\om')\delta p(z;\om')\exp(ik_xx+ik_yy-i\om't),
\ee
where $\delta p(z;\om')$ are the physical solutions satisfying the homogeneous wave equation.  When we put this into the inhomogeneous wave equation, multiply by $\rho_0\delta p^*(z;\om)\exp(-ik_xx-ik_yy+i\om t)$, and integrate over $d^3xdt$, one might think that
\bea
|A(t;\om)|\stackrel{?}{=}\int_{\ -\infty}^{\ t} d\tau\int dxdy\exp(-ik_xx -ik_yy+i\om t) \nonumber \\
\times \ \frac{1}{2\om N_0^2 \rho_0^2\sqrt{\A}} \int_{\ z_i}^{\ L} d\zeta \rho_0 \bar{S}(x,y,\zeta,\tau)\delta p^*(\zeta;\om). \label{eqn:ampdp}
\eea
Using $\delta p (\zeta;\om) \sim (\rho_0\om^2/k_\perp) \eta_z(\zeta;\om)$ (eqn.~\ref{eqn:convpol}), we see that this estimate of $|A(t;\om)|$ differs from our estimate using $\xi_z$ (eqn.~\ref{eqn:ampxiz}) by $\om^2/N_0^2$.  This leads to an underestimation of the flux in IGWs by $\sim\M^4$.

The discrepancy is due to using the incorrect inner product.  Implicit in the derivation of eqn.~\ref{eqn:ampdp} is the assumption that the $\delta p$ are orthogonal under the same inner product as the $\xi_z$, i.e.,
\be\label{eqn:badinner}
\int dz \rho_0 \partial_z \delta p(z;\om') \partial_z\delta p^*(z;\om) \stackrel{?}{=} \delta_{\om\om'} \ \rho_0^2 N_0^2.
\ee
However, one can check that the $\delta p$ are not orthogonal with respect to this inner product.\footnote{Using the properties of Hermitian operators, one can show that the $\delta p$ IGW eigenfunctions of eqn.~\ref{eqn:deltapode} are orthogonal under the inner product defined in eqn.~\ref{eqn:badinner}.  However, for the mode projection to be well defined, we must work in a complete basis, and the IGWs alone do not form a complete basis (in the convection zone).   Our resolution of this apparent inconsistency is to note that the eigenfunctions of the full non-Boussinesq wave equation {\it do} form a complete basis (this includes sound waves in addition to IGWs).  Moreover, one can show that the $\delta p$ eigenfunctions for the non-Boussinesq equations are only orthogonal under the inner product defined in eqn.~\ref{eqn:goodinner}.}  
Rather, they are orthogonal with respect to $\la a,b\ra=\int dz \rho_0^{-1} a b^*$, i.e.,
\be\label{eqn:goodinner}
\int dz \rho_0^{-1} \delta p(z;\om')\delta p^*(z;\om) = \delta_{\om\om'} \frac{\om^2}{k_\perp^2}.
\ee
Thus, if we integrate the inhomogeneous wave equation twice with respect to $z$, multiply by $\rho_0^{-1}\delta p^*(z;\om)\exp(-ik_xx-ik_yy+i\om t)$, and integrate over $d^3x dt$, we get
\bea
|A(t;\om)|=\frac{1}{2\om^3\sqrt{\A}}\int_{\ -\infty}^{\ t} d\tau\int dxdy\exp(-ik_xx-ik_yy+i\om t) \nonumber \\
\times \int_{\ z_i}^{\ L} d\zeta \rho_0^{-1}\bar{S}(x,y,\zeta,\tau)\delta p^*(\zeta;\om).
\eea
One can check that this is consistent with the calculation using $\xi_z$.

If one uses a Green's function this issue of orthogonality under different inner products becomes trivial.  Using the expansions in Sec.~\ref{sec:amplitude}, we have
\bea
 \frac{1}{\sqrt{\A}}\sum_{\om'} A(t;\om')\xi_{z,{\rm rad}}(z;\om')\exp(ik_xx+ik_yy-i\om' t) = \nonumber \\
 \int_{\ -\infty}^{\ t}d\tau \int_{\ z_i}^{\ L}d\zeta\sum_{\om'}  \frac{\xi_{z,{\rm rad}}(z;\om')\eta_z(\zeta;\om')}{N_0 k_\perp L W(\zeta)} S  \exp(-i\om'(t-\tau)), \label{eqn:gfamp}
\eea
where $z<z_i$.  Since both the left and right hand sides are in the span of $\{\xi_{z,{\rm rad}}\}_\om$, we can simply use the inner product defined by
\be
\la\xi_{z,{\rm rad}}(z;\om),\xi_{z,{\rm rad}}(z;\om')\ra = \delta_{\om\om'}.
\ee
Taking $\la\xi_{z,{\rm rad}}(z;\om),\cdot\ra$ of eqn.~\ref{eqn:gfamp}, multiplying by $\exp(-ik_xx-ik_yy+i\om t)$, and integrating in the horizontal directions, we get
\bea
A(t;\om)= \frac{1}{\sqrt{\A}}\int_{\ -\infty}^{\ t}d\tau \int dxdy\int_{\ z_i}^{\ L}d\zeta \frac{1}{N_0 k_\perp L} \frac{\eta_z(\zeta;\om)}{W(\zeta)} \nonumber \\
\times \ S(x,y,\zeta,\tau) \exp(-ik_xx-ik_yy+i\om\tau). \label{eqn:gfaamp}
\eea
This is eqn.~\ref{eqn:genamp}, which can easily be manipulated into eqns.~\ref{eqn:amps}, \ref{eqn:ampd} using the eigenfunctions.  Note that we cannot use such an inner product in the mode decomposition formalism because we need to calculate terms like $\la \delta p^*(\zeta;\om),S(x,y,\zeta,\tau)\ra$, and thus need an explicit formula for the inner product in terms of integrals over $\zeta$.

Finally, we will demonstrate that the mode projection formalism---when done correctly---and the Green's function formalism give the same result.  Specifically, we will show that eqns.~\ref{eqn:ampxiz} and \ref{eqn:gfaamp} are equivalent.  First note that $W(\zeta)$ is a constant for our wave equation.  We want to show that
\be
\frac{1}{N_0 k_\perp L W} = \frac{\rho_0\om}{2k_\perp^2}.
\ee
We can evaluate $W$ in the radiative zone, and find
\be
W=\frac{2N_0 k_\perp B_1 B_2}{\om}\sim \frac{2k_\perp}{N_0 \om L\rho_0},
\ee
where we have used eqn.~\ref{eqn:B}.  This proves that the two formulations are equivalent.

\label{lastpage}
\bibliography{radconv}

\end{document}